\documentclass[twocolumn,preprintnumbers,amsmath,amssymb,prd]{revtex4}

\usepackage{subfigure,graphicx,graphics,dcolumn,bm}
\usepackage{tabulary,tabularx}
\usepackage{epstopdf}
\newcommand{\p}{\partial}
\newcommand{\eq}{\begin{equation}}
\newcommand{\eqe}{\end{equation}}
\newcommand{\nn}{\nonumber}

\newcommand{\Mp}{M_{\mathrm{Pl}}}
\newcommand{\eqa}{\begin{eqnarray}}
\newcommand{\eqae}{\end{eqnarray}}

\newcommand{\Om}{\Omega_{{\rm fields}}}
\newcommand{\lameff}{\lambda_{{\rm eff}}}
\newcommand{\mueff}{\mu_{{\rm eff}}}

\begin{document}

\title{Quintessence dynamics with two scalar fields and mixed kinetic terms}

\author{Carsten van de Bruck} \email{c.vandebruck@shef.ac.uk}

\author{Joel M. Weller}
 \email{app07jmw@shef.ac.uk}
\affiliation{
Department of Applied Mathematics, University of Sheffield, Hounsfield Road, Sheffield S3 7RH, United Kingdom
}

\date{\today}

\begin{abstract}
The dynamical properties of a model of dark energy in which two scalar fields are coupled by a noncanonical kinetic term are studied. We show that overall the addition of the coupling has only minor effects on the dynamics of the two-field system for both potentials studied, even preserving many of the features of the assisted quintessence scenario. The coupling of the kinetic terms enlarges the regions of stability of the critical points. When the potential is of an additive form, we find the kinetic coupling has an interesting effect on the dynamics of the fields as they approach the inflationary attractor, with the result that the combined equation of state of the scalar fields can approach $-1$ during the transition from a matter dominated universe to the recent period of acceleration.
\end{abstract}


\maketitle

\section{Introduction} \label{sec:Introduction}
One goal of cosmology is to understand the origin of the observed accelerated expansion of the 
universe (see \cite{Copeland:2006wr} and references therein). To date, there are several suggestions, including the cosmological constant, slowly evolving scalar fields and modifications to Einstein's theory of general relativity. Since scalar fields are predicted by many particle physics theories, scalar field models of dark energy, such as quintessence \cite{Wetterich:1987fm, Ratra:1987rm, Martin:2008qp} or k-essence \cite{Chiba:1999ka, ArmendarizPicon:2000dh, ArmendarizPicon:2000ah} have been studied in considerable depth in the past. One requirement a satisfactory model of dark energy must fulfill is that it leads to an equation of state (EOS)
close to $w=-1$, in order to agree with current observational data. For single scalar fields, exponential potentials with slope $\lambda$ lead to a scalar field dominated universe with 
late-time accelerated expansion if $\lambda<\sqrt{2}$. In this case, the duration of the matter dominated epoch depends on the initial conditions for the scalar field.
Thus, the situation is not better than that with a cosmological constant. If, on the other hand, 
$\lambda>\sqrt{3(1+w)}$, the scalar field scales with the dominant fluid (with EOS $w$ \cite{Ferreira:1997hj, Copeland:1997et}). This would help the initial condition problem, 
but unfortunately, the scaling solution and the accelerating solution are mutually exclusive. 
The situation with inverse power-law potentials is better in the sense that a wide range of initial conditions end up with the same cosmology at late times, but in order for the theory to be consistent with observational data, the exponent has to be small. 

On the other hand, if quintessence is indeed the scenario realized in nature, the quintessence sector might turn out to be rather nontrivial. There might, for example, be several scalar fields interacting \cite{Barreiro:1999zs,Coley:1999mj,Blais:2004vt,Kim:2005ne} and/or the scalar fields might interact with matter in the universe \cite{Amendola:1999er,TocchiniValentini:2001ty, Brookfield:2007au, Bean:2008ac}. In this paper we will study interacting scalar fields as a model for dark energy and study whether this can shed some light on the issues in quintessence model building. Models with multiple scalar fields have been considered in the past: partly to study isocurvature (entropy) perturbations (e.g. \cite{Langlois:1999dw,Gordon:2000hv,Byrnes:2006fr,Langlois:2008vk}) and/or non-Gaussianity generated during 
inflation (see e.g. \cite{Alishahiha:2004eh,Rigopoulos:2005ae,Hattori:2005ac,Cai:2009hw} and 
references therein). It was also found that the cumulative effect of many scalar fields could relax the constraints on the inflationary potential \cite{Liddle:1998jc,Malik:1998gy,Calcagni:2007sb}.
These ideas have also been used in models of dark energy (assisted quintessence  \cite{Kim:2005ne,Tsujikawa:2006mw,Ohashi:2009xw}). Like the case with inflation in the 
very early universe, there is no reason why dark energy is not driven by several interacting scalar fields. In a scenario with several scalar fields one might hope to find an explanation for the coincidence problem, i.e. why dark energy dominates today and not earlier during the cosmic history. In single scalar field models this is rather difficult to achieve (see e.g. \cite{Copeland:2006wr}). 

We consider a simple extension of the standard case with canonical normalized fields by allowing for a cross-term in the kinetic energy of the scalar field, which, in the case of a homogeneous and isotropic universe, is proportional to $\dot\phi\dot\chi$. 
This is the simplest extension possible in the two-field case without changing the properties of the potential energy. 
Kinetic interactions between multiple scalar fields have been previously studied in the context of models in which the dark energy equation of state can become less than $-1$ \cite{Sur:2009jg,Chimento:2008ws}.

The paper is organized as follows: in the next section we present the model and discuss some aspects of the dynamics of the system, such as effective exponents. In Secs. \ref{sec:assist} and \ref{sec:soft} we perform a critical point analysis for two types of exponential potentials and  in Sec. \ref{sec:assist_num} we present the results of a numerical analysis. In Sec. \ref{sec:var_a} we consider the case of a varying coupling function. Our conclusions can be found in Sec. \ref{sec:discussion}.

\section{The Model} \label{sec:model}
The general action for theories which describe the dynamics of two scalar fields with noncanonical kinetic terms can be written as 
\begin{equation}
S=\int d^4x \sqrt{-g}\left[\frac{R}{2}+P(X_{mn},\phi_m)\right]
\end{equation}
where $m,n=1,2$ and
\eq
X_{mn}=-\textstyle\frac{1}{2}\left(\p_\mu\phi_m\p^\mu\phi_n\right),
\eqe
in units where $\Mp\equiv1/\sqrt{8\pi G}=1$. In this paper, we will focus on a model in which a modification to the standard two-field system, determined by the coupling $a(\phi,\chi)$, is introduced. The scalar field Lagrangian becomes 
\eq
P=X_{11}+X_{22}+a(\phi,\chi)X_{12}-V(\phi,\chi),
\eqe
where, for simplicity, we have introduced the notation $\phi_1=\phi$ and $\phi_2=\chi$. 
In the following we specialize to a flat, homogeneous and isotropic universe, where the expansion 
is described by the scale factor $R(t)$. 

If $a$ is constant, one can perform a field redefinition and diagonalise the kinetic terms so the action takes a canonical form. This is not possible in the general case, which is considered in Sec. \ref{sec:var_a}. We will  work in the field basis defined above, 
which facilitates comparison with previous work \cite{Kim:2005ne}. 
In a similar way to the analysis presented in \cite{delaMacorra:1999ff,Ng:2001hs}, solutions with constant $a$
can be considered instantaneous critical points, valid for a particular value of $a$.

\begin{figure}[htb!] 
\centering
\subfigure[]
{
\includegraphics[width=8.6cm]{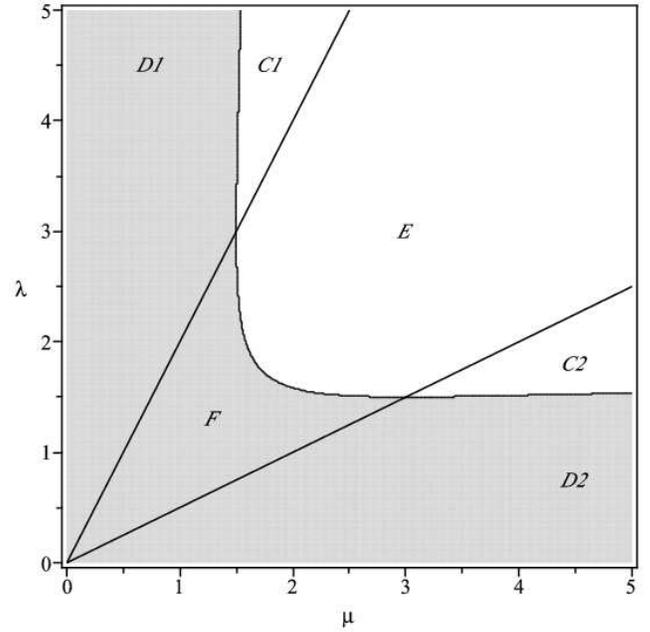}
}
\subfigure[]
{
\includegraphics[width=8.6cm]{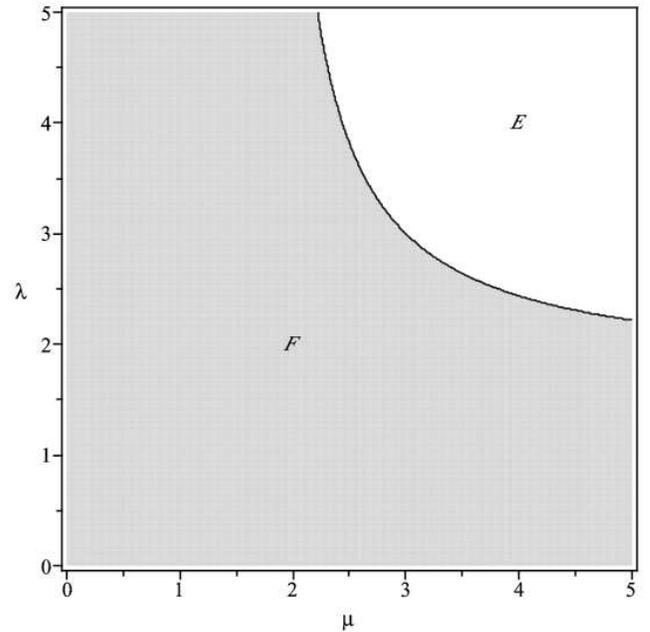}
}
\caption{Assisted case parameter space showing the stability of the critical points for (a) $a=-1$ and (b) $a=1$ with $\gamma=1$. The shaded area indicates the region with $\Om=1$. When $a$ is negative there are six stable critical points, defined by the lines $2\lambda+a\mu=0$, $2\mu+a\lambda=0$ and $\lameff^2=3\gamma$. As $a\rightarrow 0$, points E and F dominate the parameter space and when $a$ is positive, only these points are stable. }
\label{fig:assist_space}
\end{figure}

\begin{table*}[htb!]

\footnotesize

\begin{tabular*}{1\textwidth}{@{\extracolsep{\fill}} | c || c | c | c | c |}
\hline
& $x_1$ & $x_2$ & $y_1$ & $y_2$   \\
\hline
& & & &  \\ [-1em]
\hline
A & 0 & 0 & 0 & 0 \\

\hline
& & & &  \\ [-1em]
B & $-\tfrac{1}{2}x_2 a \pm \tfrac{1}{2}\sqrt{x_2^2(a^2-4)+4}  $ & $x_2$ & 0 & 0  \\

\hline
& & & &  \\ [-1em]
C1 & $-\frac{a\gamma \sqrt{6}}{4\mu}$ &
 $\frac{\gamma \sqrt{6}}{2\mu}$ & 
0 & 
$\frac{\gamma \sqrt{6}}{4\mu}\sqrt{(4-a^2)(\tfrac{2}{\gamma}-1)}$ \\

\hline
& & & &  \\ [-1em]
D1 & $ -\frac{\sqrt{6}\mu a}{3(4-a^2)} $ &
 $  \frac{2\sqrt{6}\mu}{3(4-a^2)} $ & 
 0 & 
$\frac{\sqrt{6}}{3(4-a^2)}\sqrt{(4-a^2)(6-\tfrac{3}{2}a^2-\mu^2)} $ \\

\hline
& & & &  \\ [-1em]
C2 & 
 $\frac{\gamma \sqrt{6}}{2\lambda}$ & 
 $-\frac{a\gamma \sqrt{6}}{4\lambda}$ &
$\frac{\gamma \sqrt{6}}{4\lambda}\sqrt{(4-a^2)(\tfrac{2}{\gamma}-1)}$ & 
0 \\

\hline
 & & &  & \\ [-1em]
D2 & $  \frac{2\sqrt{6}\lambda}{3(4-a^2)} $ &  
$ -\frac{\sqrt{6}\lambda a}{3(4-a^2)} $ &
$\frac{\sqrt{6}}{3(4-a^2)}\sqrt{(4-a^2)(6-\tfrac{3}{2}a^2-\lambda^2)} $ & 
0 \\

\hline
& & & &  \\ [-1em]
 E & $ \sqrt{\frac{3}{2}}\frac{\gamma}{\lambda} $ &
 $\sqrt{\frac{3}{2}}\frac{\gamma}{\mu} $ & 
$\sqrt{\frac{3}{2}}\frac{\gamma}{\mu\lambda}\sqrt{\mu(\mu+\frac{a}{2}\lambda)(\frac{2}{\gamma}-1)}$ & 
$\sqrt{\frac{3}{2}}\frac{\gamma}{\mu\lambda}\sqrt{\lambda(\lambda+\frac{a}{2}\mu)(\frac{2}{\gamma}-1)} $  \\
\hline
& & & &  \\ [-1em]
F& $ \frac{\lambda^2_{\rm eff}}{ \sqrt{6}\lambda} $  &
$ \frac{\lambda^2_{\rm eff}}{ \sqrt{6}\mu} $   &
$ \frac{\lambda_{\rm eff}}{\mu\lambda} \sqrt{ \mu \left(\mu+\frac{a}{2}\lambda\right)\left(1-\frac{\lambda^2_{\rm eff}}{6}\right) }$  & 
$ \frac{\lambda_{\rm eff}}{\mu\lambda} \sqrt{ \lambda \left(  \lambda+\frac{a}{2}\mu \right)\left(1-\frac{\lambda^2_{\rm eff}}{6}\right) }$  \\

\hline
\end{tabular*}
\caption{Critical points for the assisted potential}
\label{table:assist}
\end{table*}

In order to study scaling solutions in this unusual system we consider the scalar field system in conjunction with a perfect matter fluid of energy density $\rho$, pressure $p$ and EOS  $p=(\gamma-1)\rho$. 
The Friedmann equations are 
\eqa
3H^2 &=& \tfrac{1}{2}\dot\phi^2+\tfrac{1}{2}\dot\chi^2+\tfrac{a}{2}\dot\phi\dot\chi+V+\rho, \label{eq:friedmann} \\
\dot H &=& -\tfrac{1}{2}(\dot\phi^2+\dot\chi^2+a\dot\phi\dot\chi+\gamma\rho), \label{eq:friedmann2}
\eqae
and the 
variation of the action with respect to the fields yields 
\eqa
\ddot\phi+3H\left(\dot\phi+ \tfrac{a}{2}\dot\chi \right)+\tfrac{a}{2}\ddot\chi+V_\phi = 0, \label{eq:phi_bac}\\
\ddot\chi+3H\left(\dot\chi+ \tfrac{a}{2}\dot\phi \right)+\tfrac{a}{2}\ddot\phi+V_\chi = 0, \label{eq:chi_bac}
\eqae
where $V_\phi=\p V/\p\phi$, $V_\chi=\p V/\p\chi$ and $H = \dot R /R$.
The dot indicates a derivative with respect to time $t$. The appearance of second time derivatives of both scalar fields in each equation is characteristic of systems of this type. 
However, in the case at hand the equations can be separated for $a^2 \ne 4$, yielding
\eqa
\ddot\phi\left(1-\tfrac{a^2}{4}\right)+3H\dot\phi\left(1-\tfrac{a^2}{4}\right)+ V_\phi -\tfrac{a}{2}V_\chi= 0, \label{eq:phi_bac_ex} \\
\ddot\chi\left(1-\tfrac{a^2}{4}\right)+3H\dot\chi\left(1-\tfrac{a^2}{4}\right)+ V_\chi -\tfrac{a}{2}V_\phi= 0. \label{eq:chi_bac_ex} 
\eqae
To understand the behaviour when $a=\pm 2$, consider the reparametrization $\phi=\tfrac{1}{2}(\psi_1+\psi_2)$, $\chi=\tfrac{1}{2}(\psi_1-\psi_2)$. Substituting these expressions into Eqs. (\ref{eq:phi_bac}) and (\ref{eq:chi_bac}) gives
\eqa
\left(1+\tfrac{a}{2}\right)\left(\ddot\psi_1+3H\dot\psi_1\right)+2\frac{\p V}{\p \psi_1}= 0, \\
\left(1-\tfrac{a}{2}\right)\left(\ddot\psi_2+3H\dot\psi_2\right)+2\frac{\p V}{\p \psi_2}= 0.
\eqae
It can be seen from these equations that when $a=2$, $\psi_2$ does not play a dynamical role and its value is determined by $\psi_1$. Similarly, when $a=-2$, $\psi_1$ is not independent of $\psi_2$. Thus, when $a=\pm 2$, we are effectively dealing with a single field system.

In the following we will assume that $(4-a^2)>0$. We will perform a dynamical systems analysis, similar to that used in \cite{Copeland:1997et}, to convert the equations of motion into an autonomous system and investigate the properties of the system near the critical points. 

\subsection{Potentials}
We will study two forms of the scalar field potential, the first an additive exponential potential similar to that used in assisted inflation \cite{Liddle:1998jc,Malik:1998gy,Calcagni:2007sb} and assisted quintessence \cite{Kim:2005ne,Tsujikawa:2006mw,Ohashi:2009xw} scenarios,
\eq
V_a(\phi,\chi)=e^{-\lambda\phi}+e^{-\mu\chi}.
\eqe
This shall be referred to henceforth as the assisted potential. The attractive feature of this potential is that in the case where no matter is present, one can find a solution where the two-field system behaves like a single field with a potential less steep than either field. This allows scalar fields that are incapable of driving acceleration by themselves to co-operate with each other to sustain cosmic acceleration. It is well known that the scale factor $R(t)$ in a single field system with potential $V(\phi)=e^{-\lambda\phi}$ evolves as $R\propto t^p$, where $p=2/\lambda^2$. Applying a similar method to that used in \cite{Copeland:1999cs} to solve for $p\equiv 2/\lameff^2$ yields the effective exponent for the assisted potential in our system, 
\eq \label{eq:lameff}
\frac{1}{\lameff^2} = \frac{1}{\mu^2}+\frac{1}{\lambda^2}+\frac{a}{\mu\lambda}.
\eqe

In addition, we shall also consider a multiplicative  exponential potential, previously studied in the context of inflation in \cite{Kanti:1999vt,Copeland:1999cs,vandenHoogen:2000cf},
\eq
V_s (\phi,\chi)=e^{-\lambda\phi-\mu\chi}.
\eqe
This will be referred to as the soft potential, 
as its behaviour is similar to that of soft inflation models \cite{Berkin:1990ju} in which the presence of additional scalar fields inhibits the ability of the model to produce acceleration. Defining the effective exponent for the soft potential by $p\equiv 2/\mueff^2$ and repeating the analysis above, we find
\eq \label{eq:mueff}
\frac{1}{\mueff^2}=\frac{(1-a^2/4)}{\lambda^2+\mu^2-a\lambda\mu}.
\eqe
The effective exponents $\lameff^2$ and $\mueff^2$ can be used to simplify the untidy expressions one finds when matter is included.

\renewcommand{\tabularxcolumn}[1]{m{#1}}
\begin{table*}[htb!]

\centering
\newcolumntype{R}{>{\center}X}
\begin{tabularx}{\textwidth}[b]{ | c || R | R | c | c |  }
\hline
& Existence & Conditions for Stablity & $\Om$ & $\gamma_{{\rm fields}}$ \\
\hline
 & & &  & \\ [-1em]
\hline
  & & &  & \\ [-1.7em]
A & All $\mu$,$\lambda$,$a$ & Unstable & 0 & Undefined \\
\hline
  & & &  & \\ [-1.7em]
B & $x_2^2<(1-a^2/4)^{-1}$ & Unstable & 1 & 2 \\

\hline
  & & &  & \\ [-1.7em]
C1 & $ \mu \ge \sqrt{\frac{3\gamma}{4}(4-a^2)}$ &
$\mu < -a\lambda/2$ &
$ \frac{3\gamma}{4\mu^2}(4-a^2) $ &
$\gamma$  \\ 
\hline
 & & &  & \\ [-1.7em]
D1 & 
 $ \mu \le \sqrt{\frac{3}{2}(4-a^2)}$ &
$\mu < min\left[ -a\lambda/2,\sqrt{\frac{3\gamma}{4}(4-a^2)} \right] $&
 1 &
 $\frac{4\mu^2}{3(4-a^2)}$ \\
 
 \hline
  & & &  & \\ [-1.7em]
C2 & $ \lambda \ge \sqrt{\frac{3\gamma}{4}(4-a^2)}$ &
$\lambda < -a\mu/2$ &
$ \frac{3\gamma}{4\lambda^2}(4-a^2) $ &
$\gamma$  \\
\hline
 & & &  & \\ [-1.7em]
D2 & 
 $ \lambda \le \sqrt{\frac{3}{2}(4-a^2)}$ &
$\lambda < min\left[ -a\mu/2,\sqrt{\frac{3\gamma}{4}(4-a^2)} \right] $&
 1 &
 $\frac{4\lambda^2}{3(4-a^2)}$ \\
 
  \hline
  & & &  & \\ [-1.7em]
E & $2\lambda+a\mu >0\;\;$; $2\mu+a\lambda >0\;\;$; $ \lambda^2_{\rm eff} \ge 3\gamma $ & Stable &
 $3\gamma/ \lambda^2_{\rm eff}$ &
$ \gamma$  \\
 
  \hline
    & & &  & \\ [-1.7em]
F & $2\lambda+a\mu >0\;\;$; $2\mu+a\lambda >0\;\;$; $ 0<\lambda^2_{\rm eff}<6$  &  $ \lambda^2_{\rm eff}<3\gamma$  & 1 & $ \lambda^2_{\rm eff}/3 $ \\

\hline
\end{tabularx}
\caption{Stability of points for the assisted potential}
  \label{table:assist_stability} 
\end{table*}

\section{Assisted Potential} \label{sec:assist}

After defining new variables,
\eqa
x_1&=\frac{\dot\phi}{\sqrt{6}H}, \;\;\;\;\;\; x_2&=\frac{\dot\chi}{\sqrt{6}H}, \nn \\
 y_1&=\frac{e^{-\lambda\phi/2}}{\sqrt{3}H}, \;\;\;\;\; y_2&=\frac{e^{-\mu\chi/2}}{\sqrt{3}H},
\eqae

the autonomous system is
\eqa
x_1' &=&-3x_1+\sqrt{\frac{3}{2}}\left(1-\tfrac{a^2}{4}\right)^{-1}\left(  \lambda y_1^2  - \tfrac{a}{2}\mu y_2^2 \right)\nn \\
 &+& x_1\Xi_a, \\
x_2' &=&-3x_2+\sqrt{\frac{3}{2}}\left(1-\tfrac{a^2}{4}\right)^{-1}\left(  \mu y_2^2 - \tfrac{a}{2}\lambda y_1^2 \right)\nn \\
&+& x_2\Xi_a, \\
y_1'&=&-\sqrt{\frac{3}{2}}\lambda x_1 y_1+y_1\Xi_a, \label{eq:assist_y1}\\
y_2'&=&-\sqrt{\frac{3}{2}}\mu x_2 y_2+y_2\Xi_a,
\eqae
where $'$ indicates a derivative with respect to $N=\ln R$ and 
\eqa 
\Xi_a&=&\frac{3}{2}[  2 x_1^2 +2 x_2^2+2a x_1 x_2 \nn \\
&+& \gamma (1-x_1^2-x_2^2-a x_1 x_2-y_1^2-y_2^2) ]. \label{eq:Xia}
\eqae

We define the contribution of the fields to the energy density of the universe by
\eq
\Om=x_1^2+x_2^2+a x_1 x_2 + y_1^2+y_2^2,
\eqe
so that the constraint $0\le\Om\le 1$ arises from (\ref{eq:friedmann}).

The critical points corresponding to this system are shown in Table \ref{table:assist}. It is useful to compare these results to those in Table I in \cite{Kim:2005ne}, where the authors examine a system of two canonical scalar fields with the same potential (originally studied with nonzero curvature in \cite{Coley:1999mj}). The points with $y_1,y_2<0$ have been excluded as unphysical. It is also possible to place further constraints on the parameter space by excluding the solutions with negative values of $x_1$ and $x_2$ i.e. where the fields roll {\it up} the potential. However, as there are no stable solutions  of this kind, the unadulterated results are shown. The results of the analysis are summarised in Table \ref{table:assist_stability}, and the stability
   regions of the points are plotted in Fig. \ref{fig:assist_space}.

\begin{itemize}

\item {\bf Point A} \\
This is a trivial solution in which the scalar fields have no role to play. There are no constraints on the existence of the point so it is valid for all parameter values. As $\Om=0$ and the $y_1$ and $y_2$ values are 0 the EOS is undefined. The eigenvalues are
\eqa
&e_1&=e_2=\tfrac{3}{2}\gamma>0 \nonumber \\ & e_3&=e_4= -\tfrac{3}{2}\left(2-\gamma  \right)<0,
\eqae
so this point is a saddle.

\begin{figure*}[htp!] 
\centering 
\subfigure[]
{
\includegraphics[height=5.25cm]{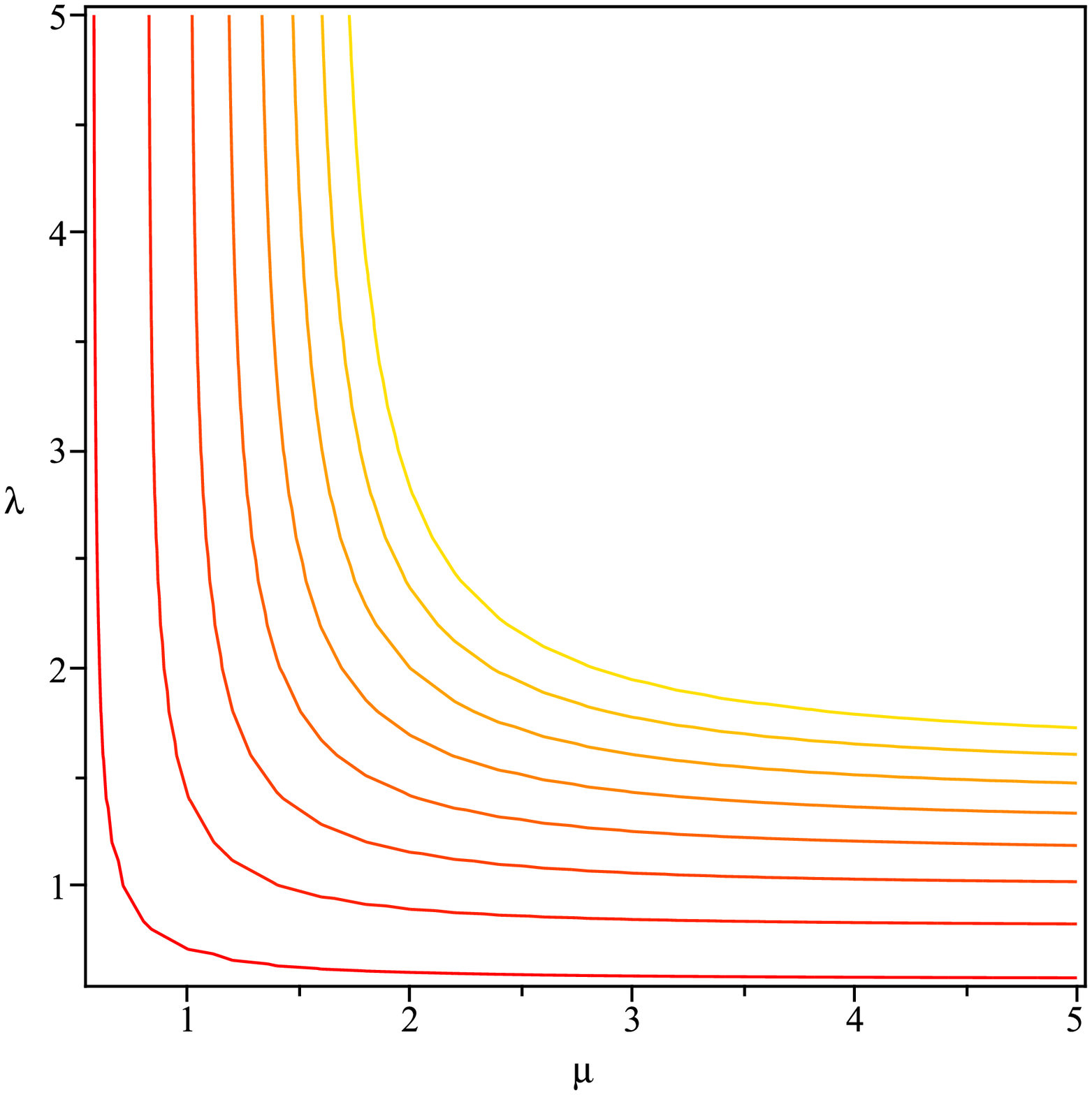}
}
\subfigure[]
{
\includegraphics[height=5.25cm]{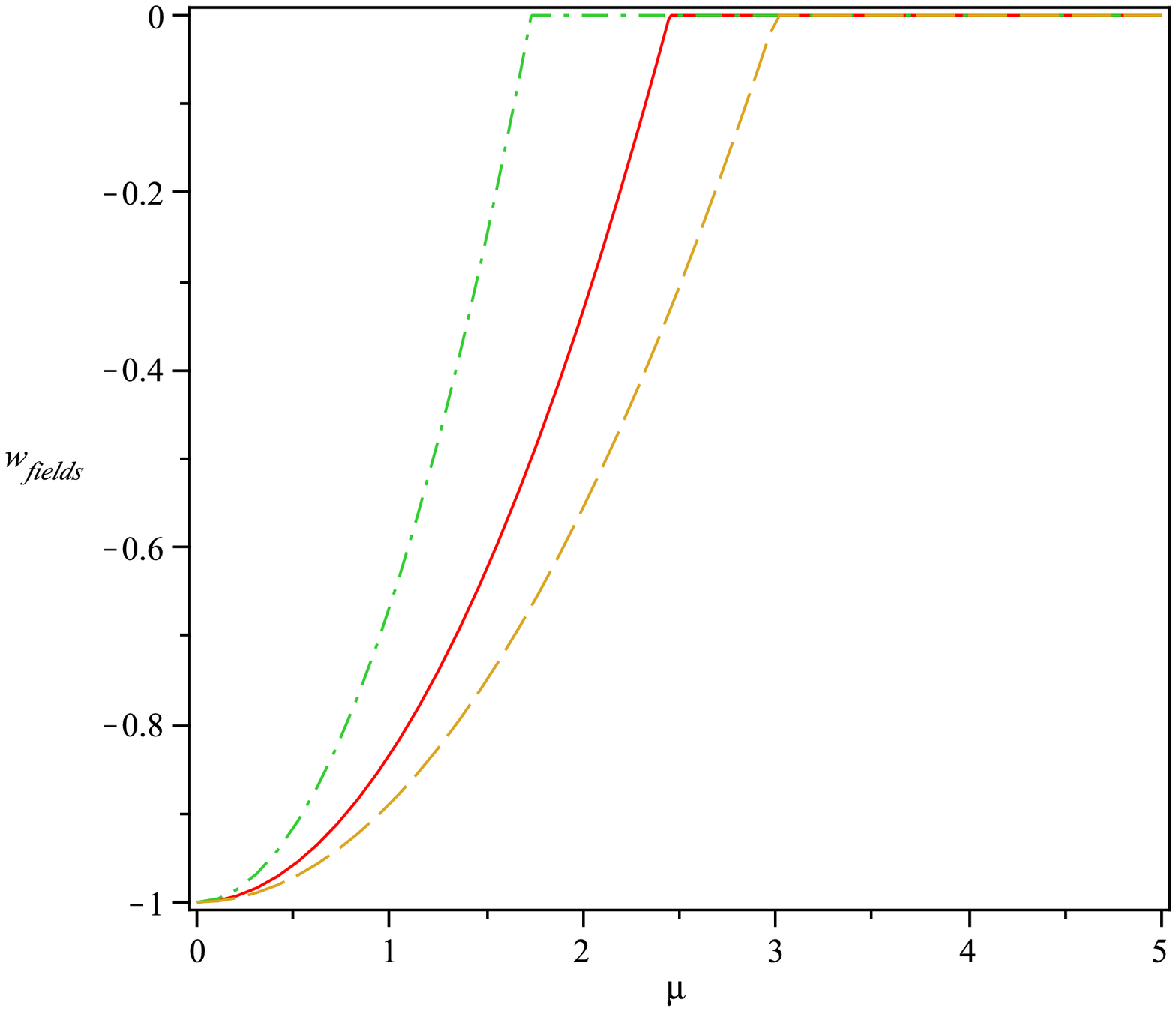}
}
\subfigure[]
{
\includegraphics[height=5.25cm]{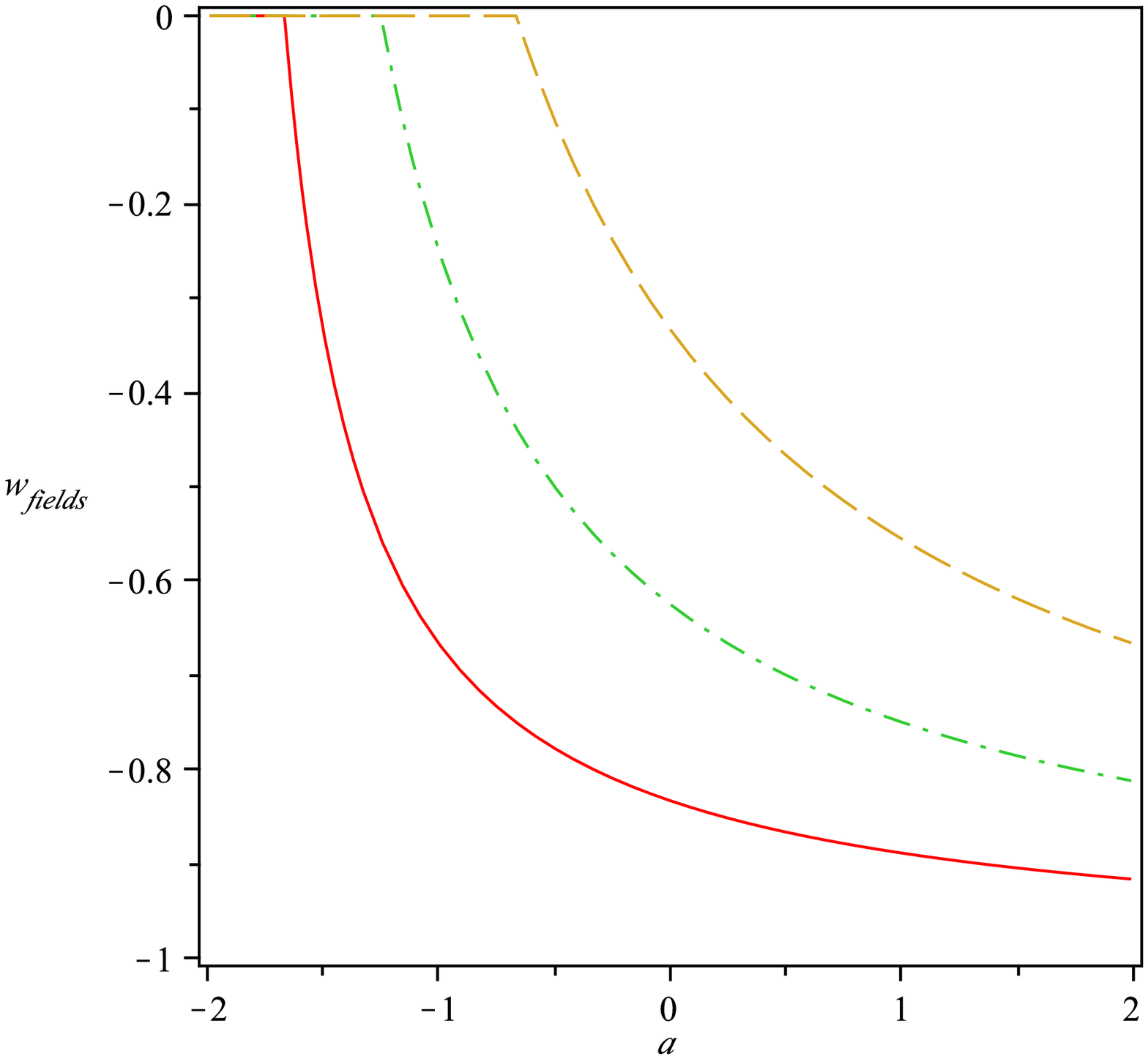}
}
\caption{The EOS for the assisted case with $\gamma=1$. (a) is a contour plot 
     showing how $w_{\rm fields}$ varies with $\mu$ and $\lambda$ for $a=0$. For large values of
     $\lambda$ and $\mu$, point E is stable so $w_{\rm fields}=0$. As $a$ varies from $-2$ to $2$,
     the contour lines spread out but otherwise do not change. (b) plots 
     $w_{\rm fields}$ along the line $\mu=\lambda$ for $a=-1$ (dot dashed), $a=0$ (solid) 
     and $a=1$ (dashed). In (c) $w_{\rm fields}$ is plotted against $a$ for 
     $\lambda=\mu=1$ (solid), $\lambda=\mu=1.5$ (dot-dash) and $\lambda=\mu=2$ (dashed).}
\label{fig:assist_w}
\end{figure*}

\item {\bf Point B} \\
The point B is a kinetically driven solution with $\Om=1$, given by,
\eqa \label{eq:pointB}
&x_1&= -\tfrac{1}{2}x_2 a \pm \tfrac{1}{2}\sqrt{x_2^2(a^2-4)+4}  \nn \\ 
&x_2& = x_2, \hspace{0.5cm} y_1 =y_2=0, 
\eqae
with $x_2$ constrained only by the requirement that $x_1,x_2\in \mathbb{R}$. The variables are real if the term under the square root is positive, giving the condition  $x_2^2<(1-a^2/4)^{-1}$. Together, the $+$ and $-$ solutions form a continuous locus of points, describing an ellipse in the $x_1,x_2$ plane (it is a circle when $a=0$). As $a^2\rightarrow 4$, the eccentricity increases and the shape rotates, anticlockwise for positive $a$ and clockwise for negative $a$. The eigenvalues are

\eqa
&e_1& = 3(2-\gamma)>0  \nn \\
&e_2& = 0 \nn \\
&e_3& = 3-\sqrt{\tfrac{3}{2}}\mu x_2 \nn \\
&e_4& = 3-\sqrt{\tfrac{3}{2}}\lambda\left(-\tfrac{1}{2}x_2 a \pm \tfrac{1}{2}\sqrt{x_2^2(a^2-4)+4}\right).
\eqae

The $\pm$ in $e_4$ corresponds to that in (\ref{eq:pointB}); $e_1$ is positive, so the point is either an unstable node or saddle depending on the values of $x_2$, $a$, $\mu$ and $\lambda$.

\item  {\bf Point C1} \\
The points C1 and D1 have the property that only one of the fields' potential energy makes a contribution to the energy density. It is interesting to note that although the points are stable for a relatively small region of the parameter space, without the coupling term in the Lagrangian the points are always unstable. The point C1 is a scaling solution in which the energy density of the fields decreases in proportion to that of the barotropic fluid. It is analogous to point 6 in \cite{Kim:2005ne}, however, the $x_1$ value is nonzero and is proportional to $-a$. The total energy density of the fields goes to 0 as $a^2\rightarrow 4$. The condition  $\Om\le1$ gives the condition for existence, $\mu\ge\sqrt{ \tfrac{3\gamma}{2}(4-a^2) }$. The eigenvalues corresponding to this point are
\eqa
&e_1&= -\tfrac{3}{2}(2-\gamma)<0, \nn \\
&e_2 &= \tfrac{3\gamma}{4\mu}(2\mu+a\lambda) \nn \\
&e_{3,4}&= -\tfrac{3(2-\gamma)}{4} \nn \\
&\;&\times\left[ 1 \pm \sqrt{ \frac{ 6\gamma^2(4-a^2)+\mu^2(2-9\gamma)}{\mu^2(2-\gamma)} }  \right].
\eqae

Using the condition for existence, one can show
\[
 6\gamma^2(4-a^2)+\mu^2(2-9\gamma) \le  \mu^2(2-\gamma),
\]
so $e_{3,4} \le  -\tfrac{3}{4}(2-\gamma)(1\pm 1) \le 0$. Thus, C1 is stable when $e_2$ is negative i.e.  $\mu<-a\lambda/2$.

\item  {\bf Point D1} \\
D1 is a scalar field-dominant solution ($\Om=1$) analogous to 5 in \cite{Kim:2005ne}, so the only existence condition comes from $y_2$. Therefore, the point exists when $\mu\le\sqrt{ \tfrac{3}{2}(4-a^2) }$. The eigenvalues for this point are
\begin{equation*}
e_1=  \frac{ \mu (2\mu + a \lambda) }{ (4-a^2) }      \hspace{0.4cm}     e_2 =  \frac{ 4\mu^2  }{ (4-a^2) } - 3\gamma \nn 
\end{equation*}
\eq
e_{3,4}=  \frac{2\mu^2}{ (4-a^2) }-3 \le 0
\eqe
where the existence condition ensures that $e_{3,4}$ are negative. If the point is to be stable both $e_1$ and $e_2$ must be negative, so the point is a stable node if
\[
\mu< min\left[ \tfrac{-a\lambda}{2}, \sqrt{\tfrac{3\gamma}{4}{(4-a^2)}} \right].
\]

\item  {\bf Points C2 and D2} \\
These points are related to C1 and D2 by the transformation $x_1 \leftrightarrow x_2 $, $y_1 \leftrightarrow y_2 $, $\lambda \leftrightarrow \mu $.  Thus, the existence and stability conditions are in terms of $\lambda$ instead of $\mu$.

\item  {\bf Point E} \\
The point E is a scaling solution analogous to 8 in \cite{Kim:2005ne}. $y_1$ and $y_2$ are square roots and $\Om \ne 1$ so there are three conditions that have to be satisfied for the point to exist. To ensure that $y_1$ and $y_2$ are real and nonzero, one has to impose $2\mu+a\lambda>0$ and $2\lambda+a\mu>0$. The third condition is given in terms of $\lameff$ by $\lameff^2 \ge 3\gamma$. The eigenvalues for this point are
\eqa 
e_{1,2}&=&-\tfrac{3}{4}(2-\gamma) \nn\\ 
&\times&\left[ 1\pm \sqrt{ \frac{24\gamma^2/\lameff^2+2-9\gamma }{(2-\gamma)} }\right], \label{eq:Eeigen12} \\
e_{3,4}&=&-\tfrac{3}{4}(2-\gamma) \nn \\
&\times&\left[ 1\pm\sqrt{ 1-\frac{8\gamma(2\mu+a\lambda)(2\lambda+a\mu)}{\lambda\mu(2-\gamma)(4-a^2)} }  \right].  \label{eq:Eeigen34}
\eqae
Using the existence condition, one finds $  \frac{24\gamma^2}{\lameff^2}+2-9\gamma \le (2-\gamma)$. Substituting this into (\ref{eq:Eeigen12}) gives $e_{1,2} \le  -\tfrac{3}{4}(2-\gamma)(1\pm 1) \le 0$. It is clear that the term under the square root in (\ref{eq:Eeigen34}) must be less than 1 if both eigenvalues are to be negative. This means the point is always stable when the conditions for existence are satisfied.

 \item  {\bf Point F} \\
 F is a scalar field-dominant solution analogous to 7 in  \cite{Kim:2005ne}. The $y_1$ and $y_2$ values are
 \eqa
 &y_1&=\frac{1}{\mu\lambda}\sqrt{ \lameff^2\mu\left(\mu+\frac{a}{2}\lambda\right)\left(1-\frac{\lameff^2}{6}   \right)  }  \nn \\ 
  &y_2&=\frac{1}{\mu\lambda}\sqrt{\lameff^2 \lambda\left(\lambda+\frac{a}{2}\mu\right)\left(1-\frac{\lameff^2}{6}   \right)  } ,
 \eqae
 so for the point to be real we require
 \eqa
 \lameff^2(2\mu+a\lambda)\left(1-\frac{\lameff^2}{6}   \right) &>&0 \nn \\
   \lameff^2(2\lambda+a\mu)\left(1-\frac{\lameff^2}{6}   \right) &>&0. \nn
 \eqae
Assuming $\left(1-\tfrac{\lameff^2}{6}   \right)<0$ and using (\ref{eq:lameff}) leads to a contradiction, so the conditions for the point to exist are,
\eq
0<\lameff^2<6, \hspace{0.3cm}  (2\mu+a\lambda)>0, \hspace{0.3cm}  (2\lambda+a\mu)>0.
\eqe
 The eigenvalues corresponding to this point are
 \eqa
 e_1&=& -3\left(1-\frac{\lameff^2}{6} \right)<0 \nn \\
 e_2 &=& \lameff^2-3\gamma \nn\\
e_{3,4} &=& -\frac{3}{2} \left(1-\frac{\lameff^2}{6}\right) \nn \\
&\times&  \left[ 1\pm \sqrt{ 1-\frac{4\lameff^2}{3\lambda\mu}\frac{(2\mu+a\lambda)(2\lambda+a\mu)}{(4-a^2)\left(1-\tfrac{\lameff^2}{6}   \right)} }  \right].
\eqae
$e_{3,4}$ are negative if the term under the square root is less than 1, which is always true as the factor in the second term under the square root is positive. Therefore, the point is stable when 
 $\lameff^2<3\gamma$.
  \end{itemize}
  In the case where $a=0$, $\lambda_{\rm eff}$ is smaller than $\lambda$ and $\mu$. This means that the points E and F are similar to the scaling and scalar field-dominant solutions of a single scalar field  \cite{Copeland:2006wr}  with a flatter potential, which naturally makes an accelerating solution more favourable (cf. assisted inflationary scenarios \cite{Liddle:1998jc}). The fact that $a$ can be both positive and negative complicates this slightly but the general principle is the same. When $a\ge0$ the conditions for existence and stability in terms of  $\lambda_{\rm eff}$ are the same as in \cite{Kim:2005ne} although it must be noted that the additive factor of $a/\lambda\mu$ means that $\lambda_{\rm eff}$ is generally smaller for given values of $\lambda$ and $\mu$. As $a\rightarrow -2$, the stability region for the scalar field dominated points shrinks. When $a<-1$ one can find combinations of fields that each have a small exponent (so considered individually they would dominate the universe) that are out of range of the $\Om=1$ attractors.

\begin{figure}
\centering
\subfigure[]
{
\includegraphics[width=8.6cm]{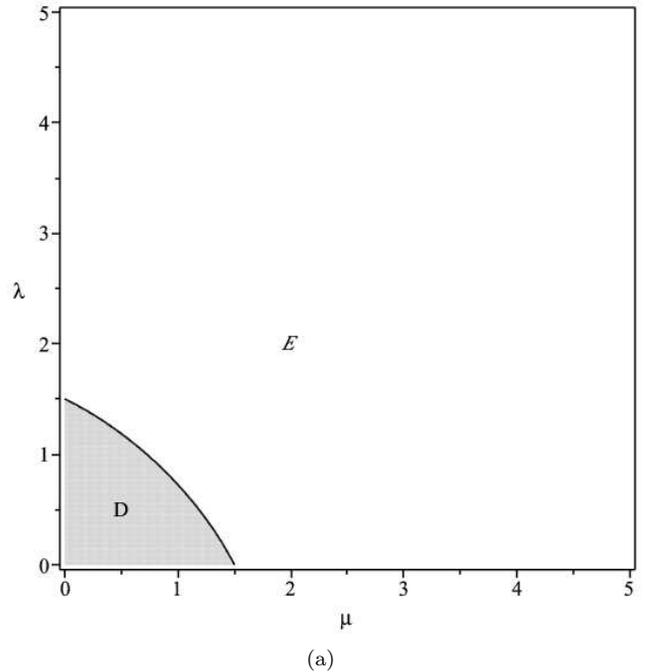}
}
\subfigure[]
{
\includegraphics[width=8.6cm]{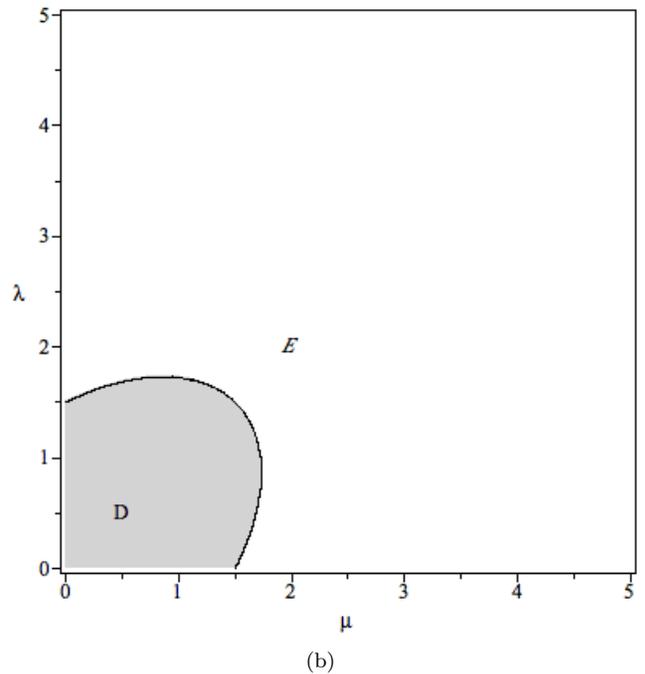}
}
\caption{Soft case parameter space showing the stability of the critical points for (a) $a=-1$ and (b) $a=1$ with $\gamma=1$. The shaded area indicates the region with $\Om=1$, where the point D in Tables \ref{table:soft} and \ref{table:soft_stability} is stable. As $a$ increases from $-2$, the shaded area grows from zero size to form a quarter circle when $a=0$, which is then squeezed along the line $\lambda=\mu$ as $a\rightarrow 2$. }
\label{fig:soft_space}
\end{figure}

\begin{table*}
\footnotesize
\centering
\begin{tabular*}{\textwidth}{@{\extracolsep{\fill}} | c || c | c | c  c|}
\hline
& $x_1$ & $x_2$ & $y$ & \\
\hline
 & & &  & \\ [-1em]
\hline
A & 0 & 0 & 0 & \\
\hline
 & & & &  \\ [-1em]
 B & $-\tfrac{1}{2}x_2 a \pm \tfrac{1}{2}\sqrt{x_2^2(a^2-4)+4}  $ & $x_2$ & 0 & \\

\hline
 & & &  & \\ [-1em]
C &$ \frac{2\sqrt{6}}{\mu^2_{\rm eff}(4-a^2)}\left[  (2\lambda-a\mu)\pm \mu\sqrt{ (4-a^2)\left(\frac{\mu^2_{\rm eff}}{6}-1  \right) } \right]$ & 
$ \frac{2\sqrt{6}}{\mu^2_{\rm eff}(4-a^2)}\left[  (2\mu-a\lambda)\pm \lambda\sqrt{ (4-a^2)\left(\frac{\mu^2_{\rm eff}}{6}-1  \right) } \right]$  &
0 & \\

\hline
 & & & &   \\ [-1em]
D &$ \sqrt{\frac{2}{3}} \frac{2\lambda-a\mu}{4-a^2}$ &
$ \sqrt{\frac{2}{3}} \frac{2\mu-a\lambda}{4-a^2}$ &
$\sqrt{ 1-\frac{\mu^2_{\rm eff}}{6}}$ &\\

\hline
 & & &   \\ [-1em]
 E & $\frac{\gamma \sqrt{6} (2\lambda-a\mu)}{\mu^2_{\rm eff}(4-a^2)}$ & 
 $\frac{\gamma \sqrt{6} (2\mu-a\lambda)}{\mu^2_{\rm eff}(4-a^2)}$ & 
$ \sqrt{\frac{3\gamma}{2}\frac{(2-\gamma)}{\mu^2_{\rm eff}} } $ & \\

\hline
\end{tabular*}
\caption{Critical points for the soft potential}
 \label{table:soft} 
\end{table*}

The observed value of the EOS of dark energy is very close to $w=-1$. A problem identified by the authors of \cite{Kim:2005ne} is that of reaching a sufficiently negative value of $w_{\rm fields}$ in a two-field system. The extra degree of freedom represented by $a$ does little to alleviate this problem when we consider the asymptotic values represented by the critical points. 
It can be seen in Fig. \ref{fig:assist_w} that although a positive value of $a$ does lead to a value of $w_{\rm fields}$ that is always smaller than the $a=0$ case, the difference is negligible in the interesting region where $w_{\rm fields} \approx -1$. 
The contour plot shows that this region is only accessible when one or more of the exponents is small, unlike the large values ($\mu,\lambda \gtrsim 1$) one would expect from fundamental theory. However, as we will discuss in 
Sec. \ref{sec:assist_num}, the kinetic coupling affects the evolution towards the critical point. In particular, during the transition from matter domination to scalar field domination, the EOS reaches a minimum near $w_{\rm fields}=-1$ before approaching the asymptotic value. 

\begin{table}
\centering
\begin{tabular}{ | c || c | c | c | c |}
\hline
& Existence &  Stablity & $\Om$ & $\gamma_{{\rm fields}}$  \\
\hline
 & & &  & \\ [-1em]
\hline
A & All $\mu$,$\lambda$,$a$ & Unstable & 0 & Undefined \\
\hline
 & & &  & \\ [-1em]
B & $x_2^2<(1-a^2/4)^{-1}$ & Unstable & 1 & 2 \\

\hline
 & & &  & \\ [-1em]
C & $ \left(\mu^2_{\rm eff}/6-1  \right)>0 $ & Unstable &1 &2 \\
 
 \hline
 & & &  & \\ [-1em] 
D & $\mu^2_{\rm eff}<6$ & 
$\mu^2_{\rm eff}<3\gamma$& 
1 & $ \mu^2_{\rm eff}/3 $\\ 
 
 \hline
 & & &  & \\ [-1em]
E & $\mueff^2 \ge 3\gamma$ & Stable &
$ 3\gamma/\mu^2_{\rm eff}$ & $\gamma$  \\
 
\hline
\end{tabular}
\caption{Stability of points for the soft potential}
 \label{table:soft_stability} 
\end{table}

When $a$ is negative, the region of the ($\mu,\lambda$)--parameter space for which $\Om=1$ [the grey shaded region in Fig. \ref{fig:assist_space}(a)] has a similar shape to that when $a$ is zero or positive but is covered by the three points D1, D2 and F.
 $w_{\rm fields}$ and the $x_1$, $x_2$, $y_1$ and $y_2$ values are continuous on the boundary. 
 In the region governed by D1, only the $\chi$ field  has a nonzero potential and, regardless of the second field, its exponent $\mu$ solely determines the character of the critical point. D2 is similarly dependent on only one field. 
These points  account for more and more of the scalar field-dominant ($\Om=1$) region of the parameter as $a\rightarrow -2$. 
The effect of this is that whenever the value of $\lambda$ is much larger than that of $\mu$ or vice versa the values of $x_1$, $y_2$, $w_{\rm fields}$ etc. are sensitive only to changes in the smaller of the two exponents.

\section{Soft potential} \label{sec:soft}

\begin{figure*}
\centering
\subfigure[]
{
\includegraphics[height=5.2cm]{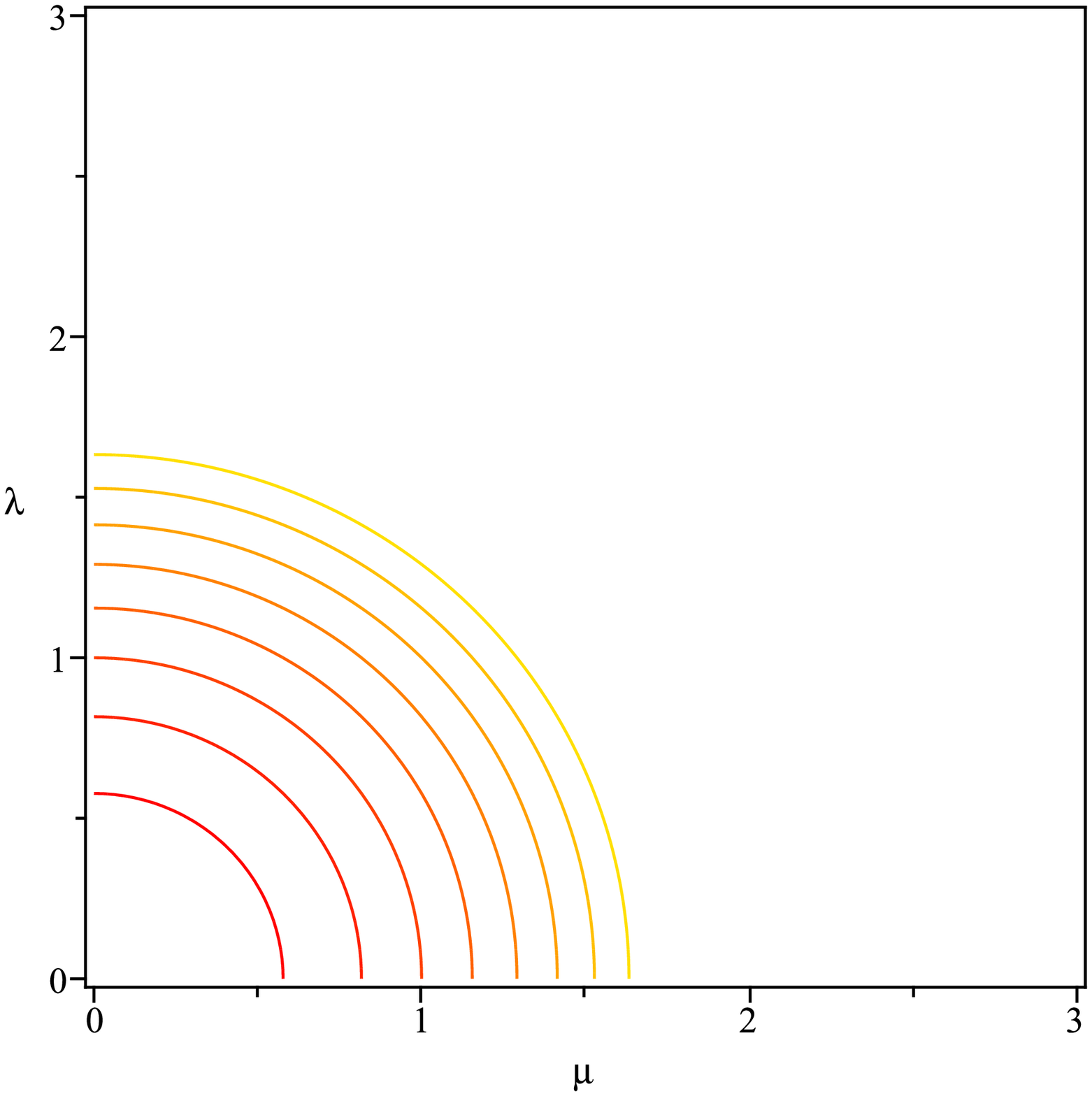}
}
\subfigure[]
{
\includegraphics[height=5.2cm]{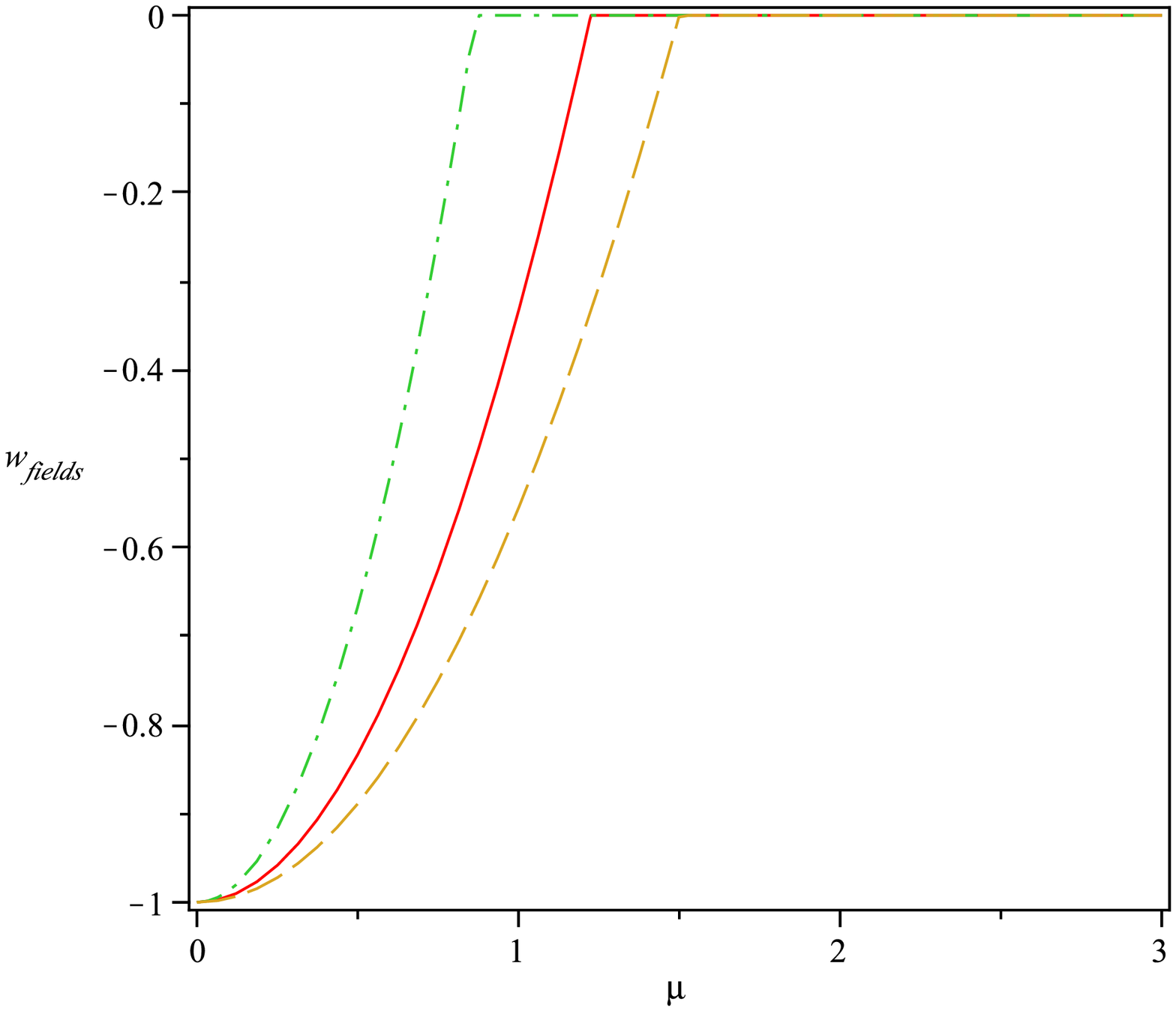}
}
\subfigure[]
{
\includegraphics[height=5.2cm]{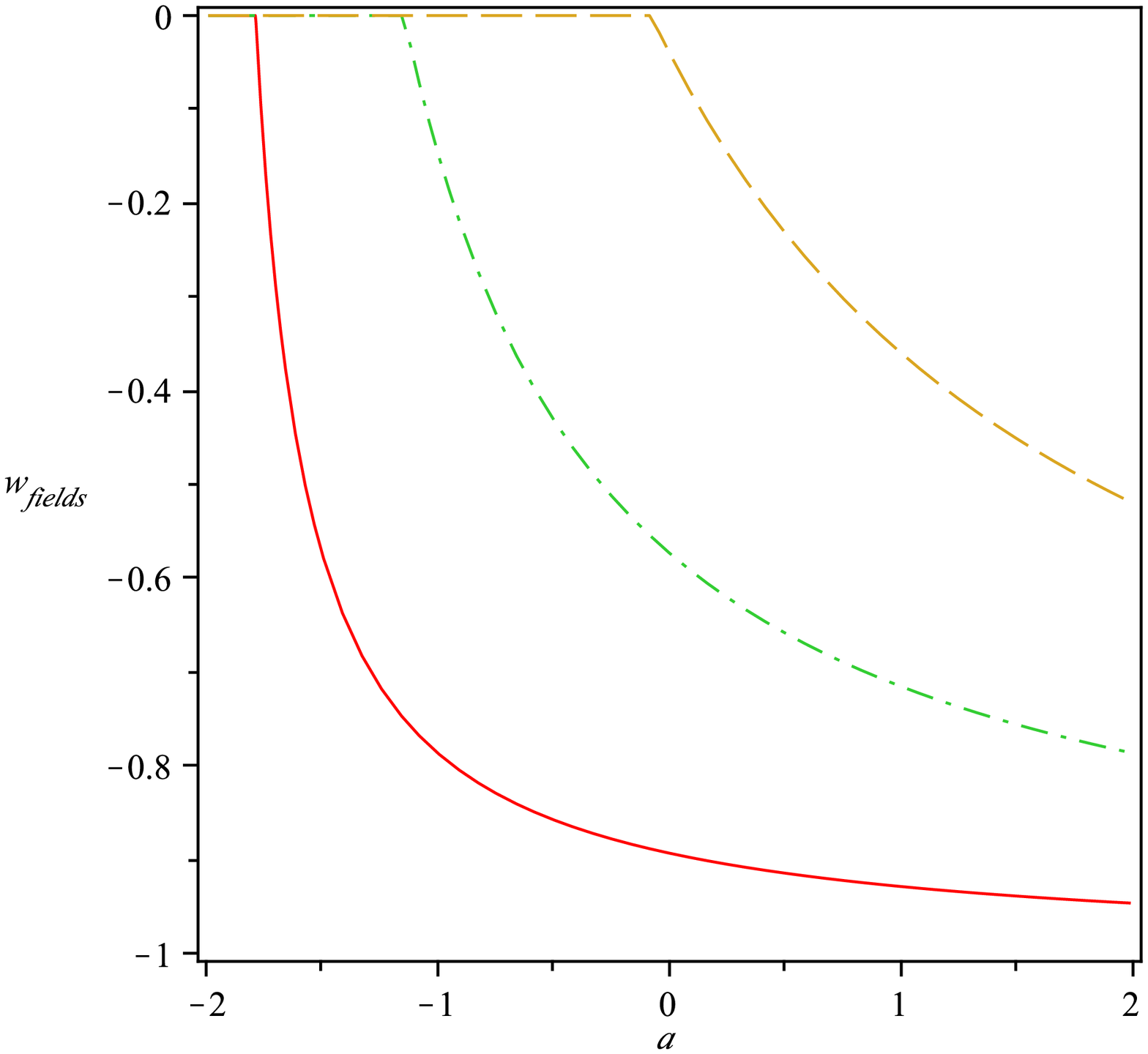}
}
\caption{The EOS for the soft case. (a) is a contour plot showing how $w_{\rm fields}$
     varies with $\mu$ and $\lambda$ for $a=0$ and $\gamma=1$. For large values of
     $\lambda$ and $\mu$, point E is 
     stable so $w_{\rm fields}=0$. (b) plots $w_{\rm fields}$ along the line 
     $\mu=\lambda$ for $a=-1$ (dot dashed), $a=0$ (solid) and $a=1$ (dashed). In (c)
     $w_{\rm fields}$ is plotted against $a$ for $\lambda=\mu=0.4$ (solid), 
     $\lambda=\mu=0.8$ (dot-dash) and $\lambda=\mu=1.2$ (dashed).}
 \label{fig:soft_w}
\end{figure*}

We define new variables for the three-dimensional soft system,
\eq
x_1=\frac{\dot\phi}{\sqrt{6}H}, \;\;\;\;\;\; x_2=\frac{\dot\chi}{\sqrt{6}H}, \;\;\;\;\;\; y=\frac{\sqrt{V_s}}{\sqrt{3}H}.
\eqe

Using these, the autonomous system is
\eqa
x_1' &=&-3x_1+\sqrt{\frac{3}{2}}\left(1-\tfrac{a^2}{4}\right)^{-1}\left(  \lambda - \tfrac{a}{2}\mu \right)y^2 \nn\\
&+& x_1\Xi_s, \\
x_2' &=&-3x_2+\sqrt{\frac{3}{2}}\left(1-\tfrac{a^2}{4}\right)^{-1}\left(  \mu - \tfrac{a}{2}\lambda  \right)y^2 \nn \\ 
&+& x_2\Xi_s, \\
y'&=&-\sqrt{\frac{3}{2}}(\lambda x_1+\mu x_2)y+ y\Xi_s, 
\eqae
where  
\eqa
\Xi_s &=&\frac{3}{2} [  2x_1^2+2x_2^2+2a x_1 x_2 \nn \\
&+&\gamma (1-x_1^2-x_2^2-a x_1 x_2-y^2) ].
\eqae

We again have the constraint $0\le\Om\le 1$ from (\ref{eq:friedmann}), but in this case the contribution of the fields to the energy density of the universe is given by
\eq
\Om=x_1^2+x_2^2+a x_1 x_2 + y^2.
\eqe

The critical points for this system are shown in Table \ref{table:soft}, again excluding unphysical values of $y$. The stability and existence conditions are investigated below, with a summary of the results in Table \ref{table:soft_stability}. The stability regions of the points are plotted in Fig. \ref{fig:soft_space}.

\begin{itemize}

\item {\bf Points A and B} \\
These points are kinetically driven and are equivalent to points A and B in the assisted case. As the potential terms are zero in each case, the stability analysis is identical to that in the previous section.

\item {\bf Point C} \\
C is actually two points, one with the $+$ sign in $x_1$ and  $x_2$ and the other with the $-$ sign. These points are also kinetically driven, with surprisingly complicated expressions for $x_1$ and $x_2$. The condition for existence is determined by the condition that the terms under the square roots must be positive.
In terms of $\mueff$, this means that
\[
\left( \mueff^2/6-1 \right) > 0,
\]
to ensure that $x_1$ and $x_2$ are real. The only nonzero eigenvalue for these points is
\eq
 e_1=3(2-\gamma)>0,
\eqe
so the points are always unstable nodes.

\item {\bf Point D} \\
D is a scalar field dominated solution somewhat similar to point F in the assisted case, as it is stable for values of the effective exponent  less than $3\gamma$. To ensure the $y$ value is real and nonzero, one requires $\mueff^2<6$. The eigenvalues are
\eq
e_{1,2}=\tfrac{1}{2}(\mueff^2-6)<0 \hspace{0.5cm} e_3 = \mueff^2-3\gamma,
\eqe
so the point is a stable node if one applies the stronger constraint $\mueff<3\gamma$.

\item {\bf Point E} \\
This is a scaling solution similar to point E in the assisted case. The square root in $y$ means that $\mueff^2>0$. The density parameter of the fields has the value $3\gamma/\mueff^2$ so the condition $\Om\le 1$ means that $\mueff^2 \ge 3\gamma$ must be satisfied if the point is to exist. The eigenvalues are
\eqa
e_1&=&-\tfrac{3}{2}<0 \nn \\
 e_{2,3}&=&-\tfrac{3(2-\gamma)}{4} \nn \\
 &\times&\left[ 1\pm \sqrt{\frac{\left(24\gamma^2/\mueff^2+2-9\gamma\right)}{(2-\gamma)}}  \right].
\eqae
The condition for existence gives $24\gamma^2/\mueff^2\le 8\gamma$, therefore $e_{2,3}\le0$ and this point is always stable.

\end{itemize}

The scalar field-dominant region of the parameter space for the soft system is much smaller and less promising than the assisted case. This can be understood by looking at the effective exponent: $\mueff$ is small only when $\mu$ and $\lambda$ are both small, so for large values of these parameters the potential is too steep to allow acceleration. There are no stable $\Om=1$ solutions with $\mu,\lambda>2$ and both of the exponents have to take small values if one is to get a solution with $w_{\rm fields}\approx-1$. As in the assisted case, negative values of $a$ lead to larger values of $w_{\rm fields}$ and vice versa, but this effect becomes less and less important when $\mu$ and $\lambda$ are small, as can be seen in Figs. \ref{fig:soft_w}(b) and \ref{fig:soft_w}(c), as near $w_{\rm fields}\approx -1$ the properties of the system  are determined by the potential terms not the kinetic terms.   
The characteristic feature of the soft potential is that the coupling in the potential inhibits the ability of the model to produce an accelerating universe. It can be seen from this analysis that the addition of the kinetic coupling exacerbates this effect.

\section{Numerical Analysis} \label{sec:assist_num}

\begin{figure*}
\centering
\includegraphics[width=1\textwidth]{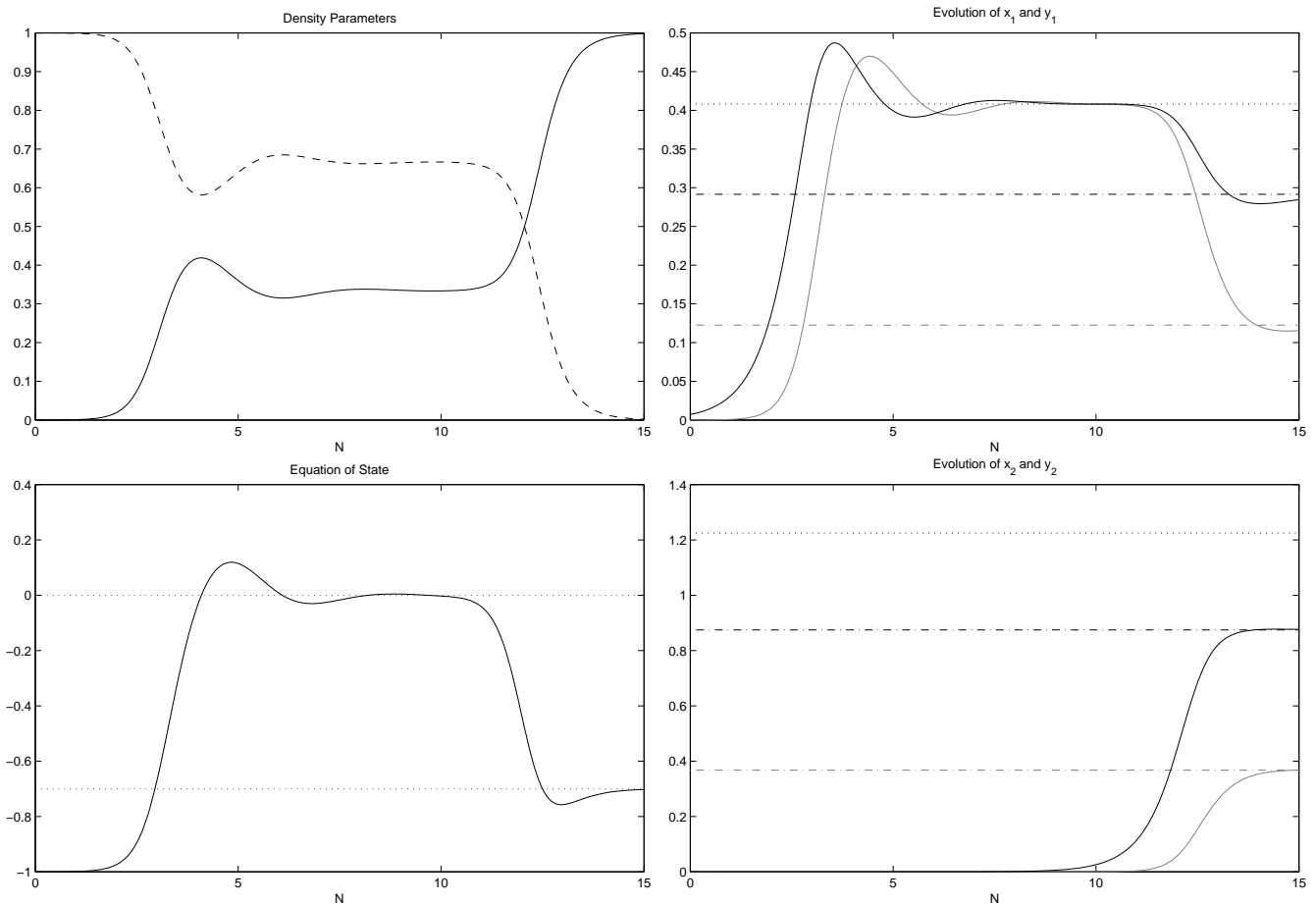}
\caption{Numerical integration for the case $a=0$, $\lambda=3$, $\mu=1$ and $\gamma=1$. As in \cite{Kim:2005ne}, the initial conditions ($\phi=1,\chi=30, \dot\chi=\dot\phi=0$) are chosen so that $\chi$ is initially negligible. The upper left panel shows $\Om$ (solid) and $\Omega_{m}$ (dashed). The lower left panel shows $w_{\rm fields}$; the upper and lower dotted lines show the expected values at critical points E and F respectively. The upper right panel shows the evolution of $x_1$ (black) and $y_1$ (grey) and the lower right panel shows the evolution of $x_2$ (black) and $y_2$ (grey). In both of these the dotted lines are the expected values at point E and the dot-dashed lines are the expected values at point F. It will be seen that the dynamical variables approach the critical point values with very little oscillation.}
\label{fig:a0}
\end{figure*}

Ideally, one would like the energy density of the scalar fields to scale with the dominant matter component before causing a period of acceleration.
Although, as we have seen in the previous section, there exist scaling solutions for this system, they are of limited phenomenological interest.
This is due to the fact that the regions for which scaling occurs and those that give rise to acceleration are mutually exclusive and the boundaries do not evolve in time.
Depending on the values of the constant parameters $\lambda$, $\mu$ and $a$, the system either does not exhibit acceleration or the scalar fields quickly dominate the universe. 
There is an interesting scenario mentioned in \cite{Kim:2005ne} which would alleviate this problem. The authors describe a situation with two scalar fields in which one field is initially negligible compared to the other.
 At first the potential of the latter is steep so the field scales with the matter component.  
When the second field becomes more important, the effective exponent decreases and the system evolves toward a scalar field dominated critical point. 
For comparison with the kinetically coupled case, this is illustrated in Fig. \ref{fig:a0}. 
If one can ignore the issue of fine-tuning, this has the advantage of providing a mechanism for late-time scalar field dominance; 
however, as mentioned previously, the value of $w_{\rm fields}$ is generally too large to compare favourably with the observational values of the EOS of dark energy.

\begin{figure*}
\centering
\includegraphics[width=1\textwidth]{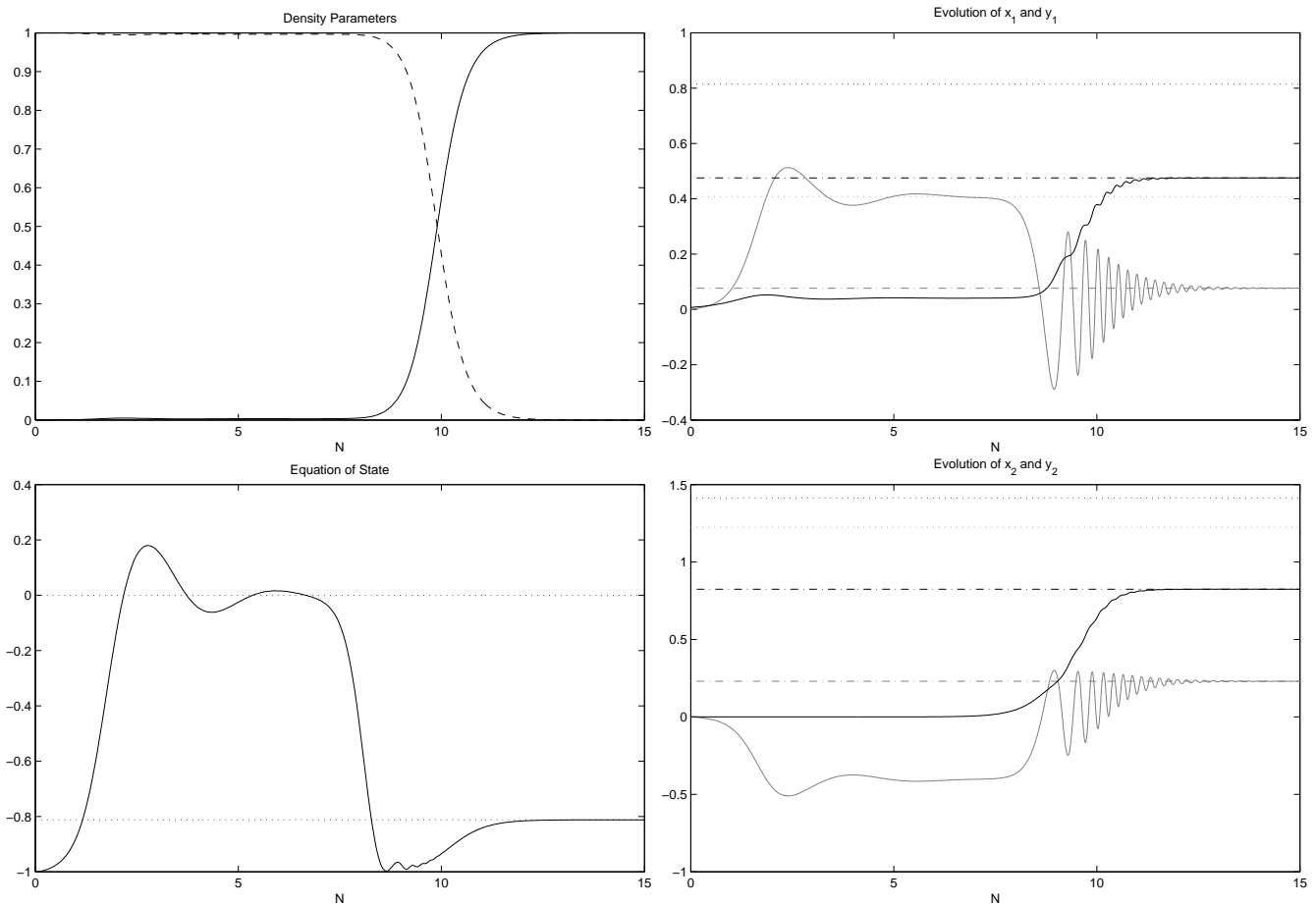}
\caption{Numerical integration for the case $a=1.99$, $\lambda=3$, $\mu=1$ and $\gamma=1$. (The quantities plotted and initial conditions are the same as those in Fig. \ref{fig:a0}.) One can see that while the fields are scaling, $\Om$ remains small and during the transition from matter to scalar field dominance, $w_{\rm fields}$ reaches $-1$.}
\label{fig:a199}
\end{figure*}

If one imposes similar initial conditions on the two-field system with mixed kinetic terms, the results are more promising. As one can see in Fig. \ref{fig:a199} (which differs from Fig. \ref{fig:a0} only in the value of $a$), there are two major differences: the energy density of the fields is negligible until the end of the matter dominated era and, as the transition occurs, the EOS of the fields drops to $-1$ before reaching the value given in Table. \ref{table:assist_stability}. The effect becomes more pronounced as $a\rightarrow 2$. A similar phenomenon in which the EOS reaches a minimum before approaching the final value has been observed recently for other scalar field models \cite{Ohashi:2009xw}, however the effect is more dramatic and prolonged in our case.

This behaviour can be understood by examining the evolution of $x_1$, $x_2$, $y_1$ and $y_2$. In the $a=0$ case and with the $\chi$ field initially negligible, the kinetic ($x_2$) and potential ($y_2$) terms of the $\chi$ field do not play a role until eventually $y_2$ increases. In the $a=1.99$ case, $\exp (-\mu\chi_{\rm ini}) $ is small so $V_\chi \ll V_\phi$. In this case, the equations of motions for both fields  [see (\ref{eq:phi_bac_ex}) and (\ref{eq:chi_bac_ex})]
reduce to 
\eqa
\frac{1}{a^3}\left(\dot\chi a^3  \right)\dot{} \approx \frac{V_{\phi}}{\left( 1-a^2 / 4  \right)} \nn \\
\frac{1}{a^3}\left(\dot\phi a^3  \right)\dot{} \approx -\frac{V_{\phi}}{\left( 1-a^2 / 4  \right)} \nn
\eqae
Integrating these equations, with the initial conditions $\dot\phi=\dot\chi=0$, we find 
$\dot\chi = -\dot\phi$ and so $x_2\approx -x_1$, as can be seen in Fig. \ref{fig:a199}. The factor of $(1-a^2 / 4)$ in the denominators of the equations above increases the values of $\dot\phi$ and $\dot\chi$ compared to the $a=0$ case so when the system approaches the scalar field-dominant solution they overshoot the critical point and oscillate around it. As the term involving the coupling ($a x_1 x_2$) is significantly large and negative, $w_{\rm fields}$ is decreased during this period.

Assuming $x_2=-x_1$,  $y_1, y_2\ll1$ and $a\approx 2$ means that the function $\Xi_a$ (defined in (\ref{eq:Xia})) simplifies to $\Xi_a\approx \tfrac{3}{2}\gamma$. This means that (\ref{eq:assist_y1}) becomes
\eq
y_1' \approx \frac{3}{2}\gamma y_1 \left( 1 - \sqrt{\frac{2}{3}}\frac{\lambda}{\gamma}x_1  \right).
\eqe
As in the $a=0$ case, the system first approaches the scaling point E, which means $x_1\rightarrow \sqrt{\tfrac{3}{2}}\gamma/\lambda$. Thus, the evolution of $y_1$ is quickly halted and, until the $\chi$-field becomes significant, the fields only occupy a fraction of the energy density of the universe. Effects of this type are not seen in the case of the soft potential as the condition  $V_\chi \ll V_\phi$ corresponds to $\mu \gg \lambda$, which is not realised in the region where the $\Om=1$ attractor is stable.

\begin{figure*}

\centering
\makebox[\textwidth][c]{
\subfigure[]
{
\includegraphics[height=12.5cm,width=0.57\textwidth]{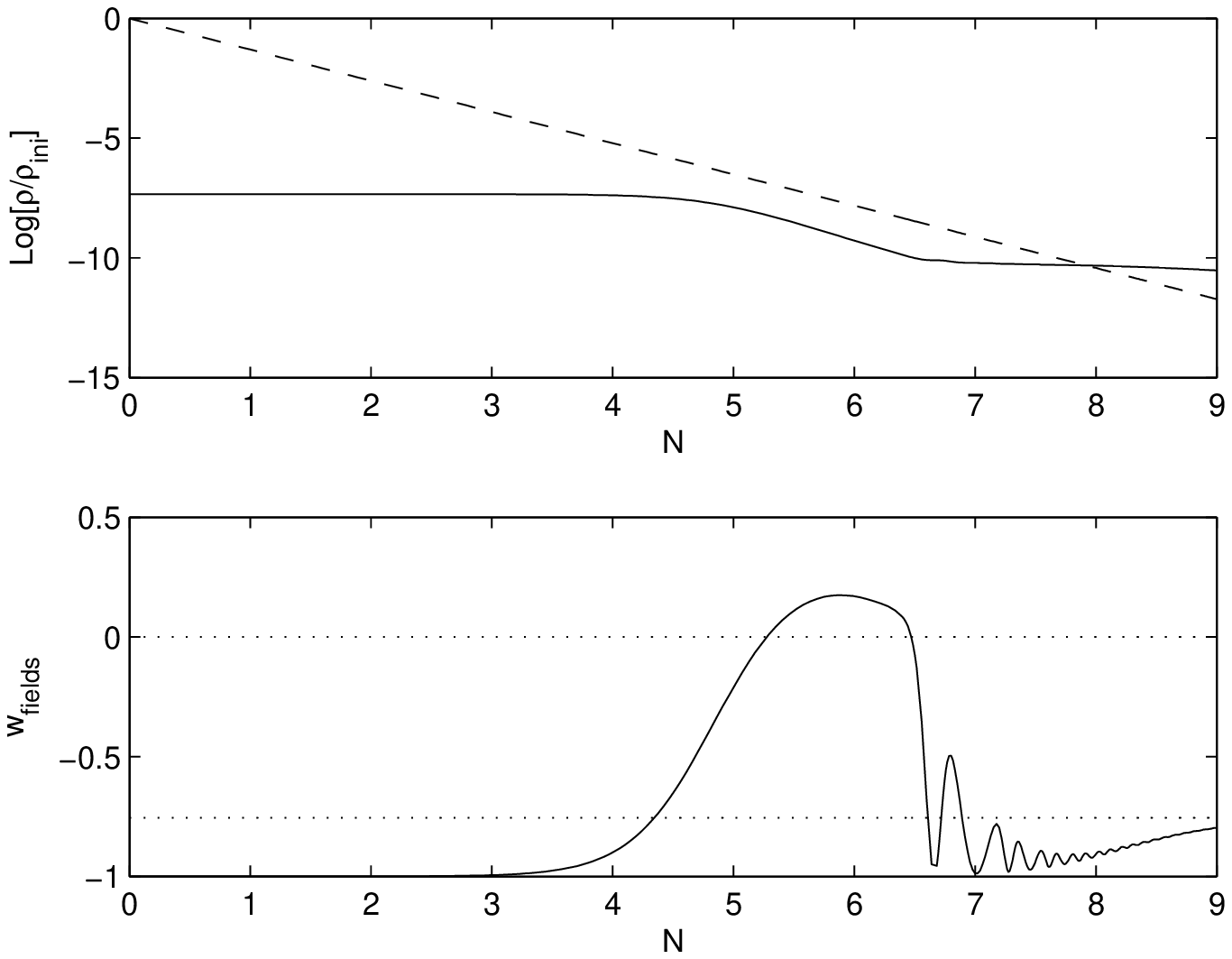}
}
\hspace{-1.3cm}
\subfigure[]
{
\includegraphics[height=12.5cm,width=0.57\textwidth]{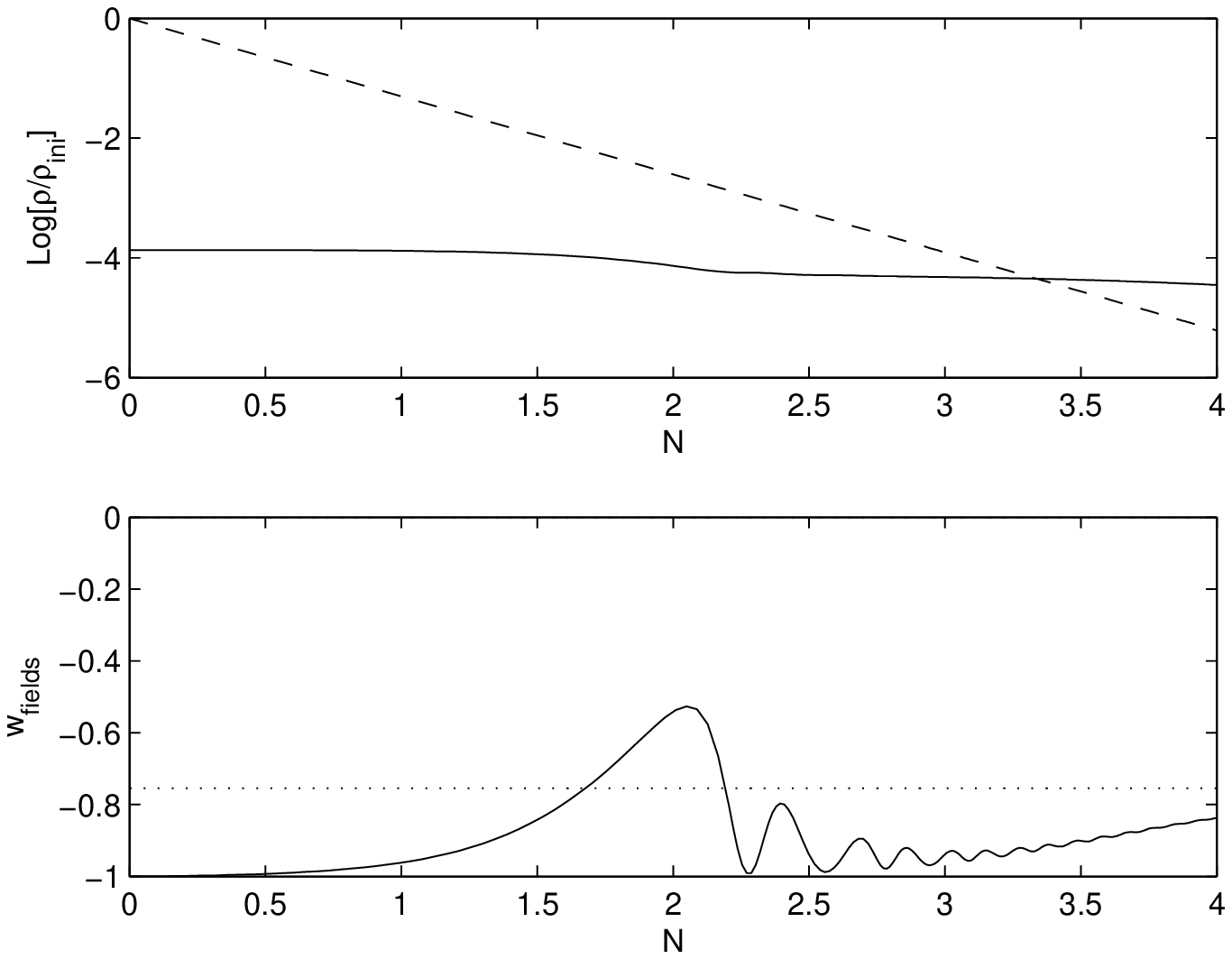}
}
}
\caption{Two trajectories for different initial conditions  ($\phi_{\rm ini}=\chi_{\rm ini}$), for $\lambda=3$, $\mu=1$ and $\gamma=1$. The upper plots show the evolution of the energy densities of the fields (solid) and matter (dashed) and the lower plots show the scale factor, $w_{\rm fields}$.}
 \label{fig:num}
 
\end{figure*}

This feedback effect occurs whenever $a$ is close to 2 and $V_\chi \ll V_\phi$ (or $V_\phi \ll V_\chi$) so is affected both by the initial conditions and the relative sizes of $\lambda$ and $\mu$. 
It is noteworthy that one does not actually require the initial conditions 
	 of the two fields to be different to see the scaling
	 effect and the decrease in $w_{\rm{fields}}$ (see Fig. \ref{fig:num}).
If the initial contribution of the scalar fields to energy density is small compared to that of matter, a small difference in the values of $\mu$ and $\lambda$ will yield the condition $x_1\approx-x_2$ that
pushes the EOS to $-1$. 
When the system is initially close to the critical point there is no scaling regime, but as long as $V_\chi \ll V_\phi$ and $a\approx 2$, one still observes the effect on $w_{\rm fields}$. However, although the scalar field dominated solution is an attractor, the question of when the scalar fields become cosmologically important is dependent the initial conditions.

We can gain further insight into this behaviour (in the case of constant $a$) by performing a field redefinition to diagonalise the kinetic terms,
\eqa
\varphi_1 &\equiv& \sqrt{\tfrac{1}{2}\left( 1-\tfrac{a}{2} \right)}\left( \phi-\chi \right), \\
\varphi_2 &\equiv& \sqrt{\tfrac{1}{2}\left( 1+\tfrac{a}{2} \right)}\left( \phi+\chi \right), 
\eqae
so that the Lagrangian for the scalar field system with the assisted potential becomes,
\eqa
\mathcal{L}=-\tfrac{1}{2}\left(\p_\mu\varphi_1\p^\mu\varphi_1\right)-\tfrac{1}{2}\left(\p_\mu\varphi_2\p^\mu\varphi_2\right) \nn\\ 
-e^{-\lambda(A\varphi_1+B\varphi_2)}-e^{-\mu(B\varphi_2+A\varphi_1)},
\eqae
with $A=1/\sqrt{2-a}$ and $B=1/\sqrt{2+a}$. As $a \rightarrow 2$, $B \rightarrow 1/2$ and $\varphi_1 \rightarrow 0$ [note that the combination $A\varphi_1=\tfrac{1}{2}(\phi-\chi)$ remains finite]. Thus the Lagrangian can be written
\eq
\mathcal{L}=-\tfrac{1}{2}\left(\p_\mu\varphi_2\p^\mu\varphi_2\right)-V_1 e^{\lambda\varphi_2/2} -V_2 e^{-\mu\varphi_2/2},
\eqe 
with $V_1=e^{-\lambda A\varphi_1}$ and  $V_2=e^{\mu A\varphi_1}$ constant. This is similar to the case in \cite{Barreiro:1999zs}, where a double exponential potential is used to give a scaling regime followed by acceleration driven by the scalar field.  

\section{Field Dependent Coupling}
\label{sec:var_a}

\begin{figure}[t]
\includegraphics[width=8.6cm,height=10cm]{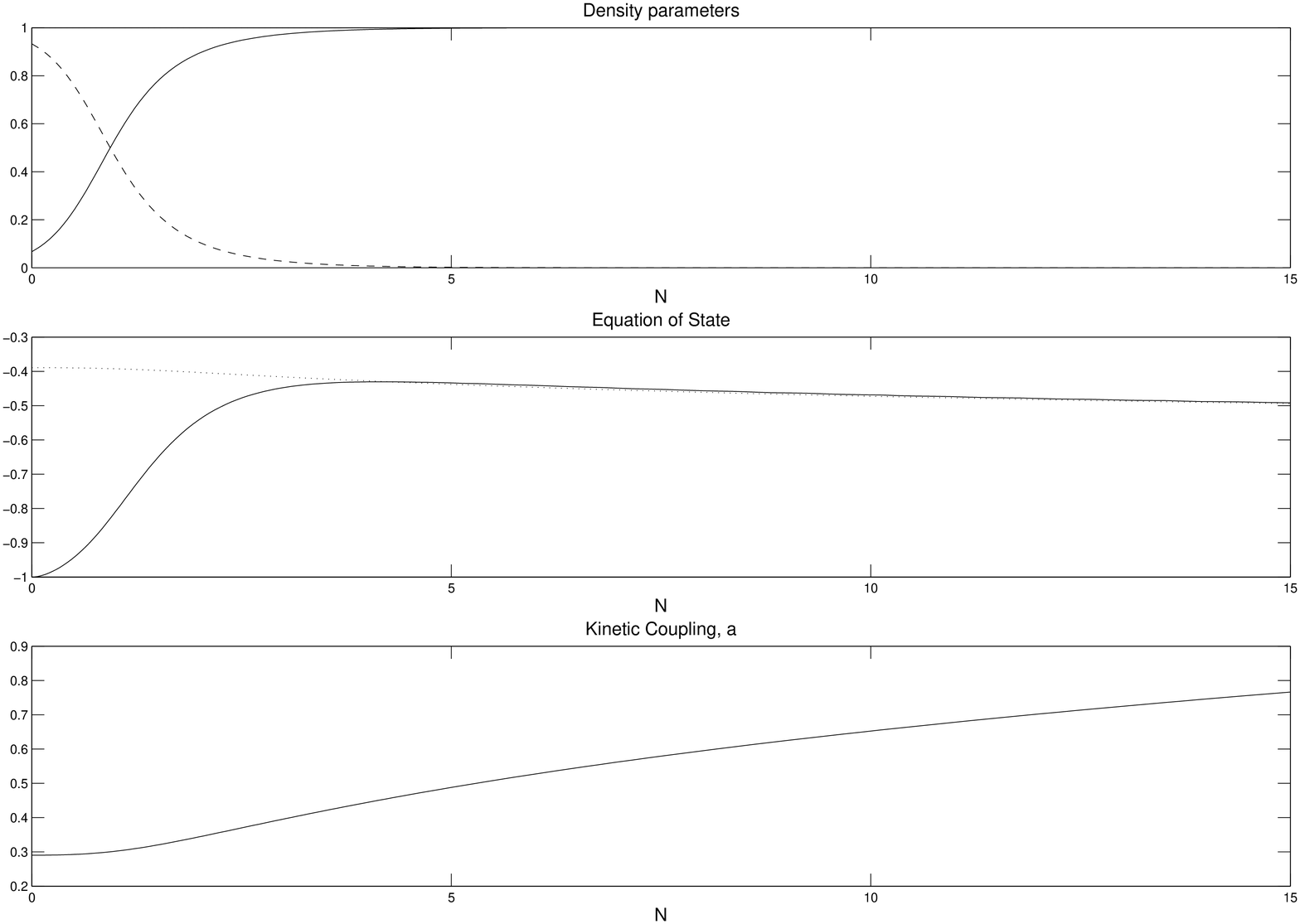}
\caption{A slowly varying coupling function, modelled by $a(\phi,\chi)=\log_{10}\left(\tfrac{1}{2}(\phi+\chi)+5\right)-\tfrac{9}{20}$, with $\phi_{\rm{ini}}=\chi_{\rm{ini}}$, $\lambda=2$, $\mu=2.1$ and $\gamma=1$. The upper panel shows $\Om$ (solid) and $\Omega_{m}$ (dashed) and the middle panel shows the EOS (solid) and the value of $w_{\rm fields}$ at the critical point, $(w_{\rm fields})_F$ (dotted). The lower panel shows  the evolution of $a$. The system initially heads toward the field-dominant point F. As $a$ increases, $w_{\rm fields}$ tracks the change in  $(w_{\rm fields})_F$.
}
\label{fig:loga}

\end{figure}

\begin{figure}[t]
\includegraphics[width=8.6cm,height=10cm]{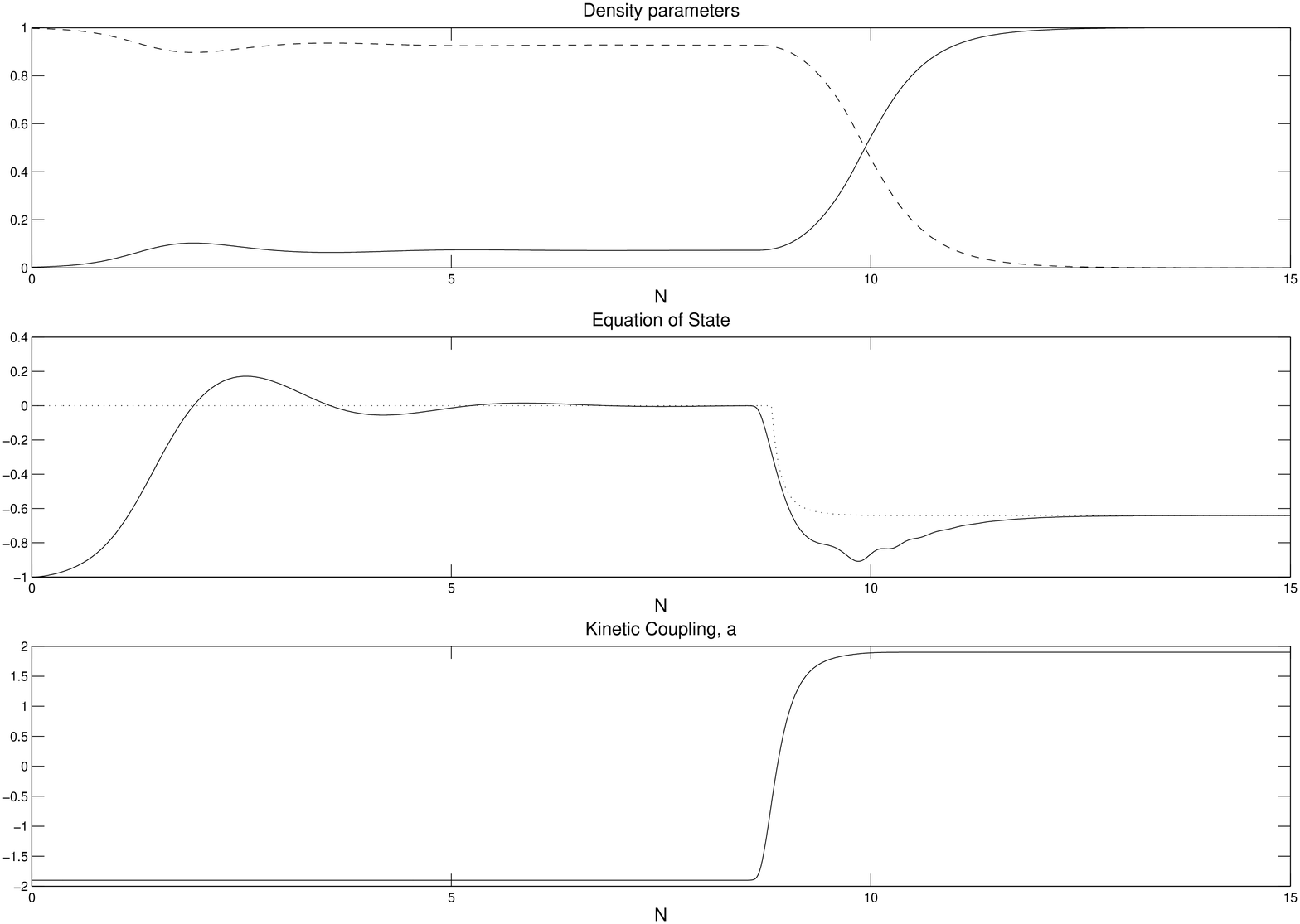}
\caption{A rapid shift in $a$, modelled by $a(\phi,\chi)=1.9\times\tanh\left(8[(\phi+\chi)-25] \right)$, with $\phi_{\rm{ini}}=\chi_{\rm{ini}}$, $\lambda=2$, $\mu=2.1$ and $\gamma=1$. The quantities plotted are the same as those in Fig. \ref{fig:loga}, except that when point F is unstable, the dotted line in the second panel shows $(w_{\rm fields})_E=0$.  Initially the point E is stable and the system exhibits scaling. After $a$ changes rapidly the existence conditions for E are no longer satisfied and the system evolves toward the field dominant point F.}
\label{fig:tanh}
\end{figure}

In the previous analysis, we assumed that the kinetic coupling, $a$, is constant in time and independent of the scalar fields. In this section we relax this assumption and extend the analysis to include $a=a(\phi,\chi)$ for the assisted potential. The Friedmann equations (\ref{eq:friedmann}) and (\ref{eq:friedmann2}) are unaffected but the equations of motion (\ref{eq:phi_bac_ex}) and (\ref{eq:chi_bac_ex}) have additional terms on the right-hand side,
\eqa
\ddot\phi+3H\dot\phi &+& \left(1-\tfrac{a^2}{4}\right)^{-1}\left(V_\phi -\tfrac{a}{2}V_\chi\right) \nn \\
&=& \left(1-\tfrac{a^2}{4}\right)^{-1}\left( \tfrac{1}{4}a a_\phi \dot\phi^2-\tfrac{1}{2}a_\chi \dot\chi^2 \right), \label{eq:phi_bac_ex_a} \\
 \ddot\chi+3H\dot\chi &+& \left(1-\tfrac{a^2}{4}\right)^{-1}\left(V_\chi -\tfrac{a}{2}V_\phi\right) \nn \\
&=& \left(1-\tfrac{a^2}{4}\right)^{-1}\left( \tfrac{1}{4}a a_\chi \dot\chi^2-\tfrac{1}{2}a_\phi \dot\phi^2 \right) ,\label{eq:chi_bac_ex_a} 
\eqae
where $a_\phi$ and $a_\chi$ are the partial derivatives of $a$ with respect to the fields $\phi$ and $\chi$ respectively.
Our previous analysis is valid for two cases of particular interest: a slowly varying coupling and a rapid change between approximately constant values of $a$. In the former case, the effect of the extra terms on the dynamics of the fields is very slight and we can consider the asymptotic solutions discussed in Sec. \ref{sec:assist} as instantaneous critical points (cf. \cite{delaMacorra:1999ff,Ng:2001hs}), valid for particular values of $a(\phi,\chi)$. The system will evolve toward one of the stable critical points as before; as $a$ varies only slowly, the solution will track the variation in $a$. This can be seen in Fig. \ref{fig:loga}, where the solution tracks the point F as $a$ slowly increases to larger values.

When the rate of change of $a$ is faster than the typical response time of the system, in order to understand the behaviour one would typically have to study the effect of the terms in $a_\phi$ and $a_\chi$ in (\ref{eq:phi_bac_ex_a}) and (\ref{eq:chi_bac_ex_a}) for a particular model. An interesting exception to this is when the timescale during which $a$ changes is short. An example of this is illustrated in Fig. \ref{fig:tanh}, where we have chosen $a(\phi,\chi)\propto \tanh(C(\phi+\chi)+{\rm constant})$. This allows us to generate a rapid change in the coupling $a$. At first, $a$ is approximately $-1.9$ so the system exhibits scaling. When the system has reached the scaling point (E in this case), $w_{\rm fields}=0$ and there can be no acceleration. The scaling regime is abruptly ended when $a\rightarrow+1.9$. Point E no longer exists and F is stable instead, so the scalar fields will drive the acceleration of the universe.

To conclude, the results of the phase space analysis in the case of constant $a$ are useful when $a$ is a 
slowly evolving function as well as the case of a spontaneous change in the value of $a$ (followed by a period where the coupling is approximately constant).

\section{Discussion} \label{sec:discussion}
Although the use of scalar fields to model the acceleration of the universe could well be described as standard practice in cosmology, the fundamental physics that underpins these phenomenological models is by no means clear. It is wise, therefore, to consider the possibility that the observed acceleration of the universe may have more exotic origins. The recent focus on scalar fields with noncanonical kinetic terms derived from string-inspired models has shown that models that deviate from the standard lore are capable of producing fruitful results that display qualitatively different behaviour. In this work we have considered a model of dark energy in which the kinetic terms of two scalar fields are coupled by a term in the Lagrangian of the form $a(\p_\alpha\phi\p^\alpha\chi)$. This allowed us to study the simplest extension of the standard case without modifying the potential energy. 
After analysing the phase plane structure of the model for two different potentials, we found that the basic features of the uncoupled case were preserved and the most important effect of the extra degree of freedom was to change the range of parameters that lead to the different types of solutions. In particular, there exist stable points corresponding to scalar field-dominant or scaling solutions for every value of $\lambda$, $\mu$ and $a$ considered.

In the case of the assisted potential we found that for negative $a$, critical points C1, D1, C2 and D2 that existed but were unstable in the case without the kinetic coupling became stable. 
This means that when $a$ goes from positive to negative values, there are three different regions in the $\lambda,\mu$-parameter space that give rise to accelerated solutions, each with a different value of the EOS, $w_{\rm fields}$. 
This is less interesting from a phenomenological point of view, however, as except when the values of the potential exponents are very small, negative $a$ values have $w_{\rm fields}$ larger than the observed dark energy EOS. 
In the case of positive $a$, one finds the EOS to be closer to $-1$ than that in the model with $a=0$, though in the region of interest ($w_{\rm fields}\approx-1$) the effect of the kinetic coupling is extremely slight. The soft potential evinced a similar dependence on $a$, though it is of less interest as a cosmological model  as accelerated behaviour only occurs with small values of the exponents in all cases.

Although models with exponential potentials exhibit both scaling and scalar field-dominant solutions, it is difficult to solve the coincidence problem without fine-tuning the initial conditions or allowing the parameters to vary in time. In Sec. \ref{sec:assist_num} we discuss how the kinetic coupling affects the dynamics of the fields as they evolve towards the critical points. When $a$ is close to $2$ one finds that the value of $w_{\rm fields}$ becomes very close to $-1$ for a brief period before the scalar fields dominate. This is a partial resolution of the perennial problem of the EOS being too large.

This analysis is revealing as the addition of a cross-kinetic term has only a relatively minor effect on the dynamics of the two-field system for both cases studied, even preserving many of the features of the assisted quintessence scenario. In the case of the assisted potential, the kinetic coupling has an interesting effect on the dynamics of the fields as they approach the stable solution, with the result that the EOS of the scalar fields can approach $-1$ during the transition from a matter dominated universe to the recent period of acceleration, something that does not occur in assisted quintessence or the single field case. A natural extension to this work is to consider the case when the scalar fields are coupled directly to the matter fluid by a similar mechanism to that described in \cite{TocchiniValentini:2001ty}. In this case, one would expect the value and stability ranges of the critical points to change dramatically due to the extra degree(s) of freedom, but it would be interesting to see whether the additive potential still allows large values of the potential exponents. The work could also be extended by treating $\lambda$ and $\mu$ as dynamical variables in a manner similar to \cite{Ng:2001hs}, in which case it may be possible to obtain solutions with a large range of initial conditions that track the background solution and later dominate.

As we have discussed, our results are also applicable in the case of a slowly varying coupling function $a(\phi,\chi)$  as well 
as the case in which the coupling undergoes a rapid change but is constant (or evolving very slowly) after the transition. 

\begin{acknowledgments}
JMW is supported by EPSRC and CvdB is partly supported by STFC. We would like to thank Rekha Jain for useful discussions.
\end{acknowledgments}

\bibliography{/Users/mozacz/Documents/Reports/refs.bib}

\end{document}